\documentclass[10pt,a4paper]{article}

\usepackage[a4paper,margin=2.2cm]{geometry}
\usepackage[T1]{fontenc}
\usepackage{lmodern}
\newcommand{\figfont}{\sffamily}
\usepackage{amsmath,amssymb,amsthm}
\usepackage{mathtools}
\usepackage{bm}
\usepackage{graphicx}
\usepackage{xstring}
\let\OriginalIncludeGraphics\includegraphics
\renewcommand{\includegraphics}[2][]{%
  \IfEndWith{#2}{.tif}{%
    \StrBefore{#2}{.tif}[\TifGraphicsBase]%
    \OriginalIncludeGraphics[#1]{\TifGraphicsBase.pdf}%
  }{%
    \IfEndWith{#2}{.tiff}{%
      \StrBefore{#2}{.tiff}[\TifGraphicsBase]%
      \OriginalIncludeGraphics[#1]{\TifGraphicsBase.pdf}%
    }{%
      \OriginalIncludeGraphics[#1]{#2}%
    }%
  }%
}
\usepackage{tikz}
\usetikzlibrary{positioning,arrows.meta,calc}
\usepackage{booktabs}
\usepackage{multirow}
\usepackage{threeparttable}
\usepackage{array}
\usepackage{tabularx}
\usepackage{makecell}
\usepackage{caption}
\usepackage{subcaption}
\usepackage{xcolor}
\usepackage{pdflscape}
\usepackage{placeins}
\usepackage{authblk}
\usepackage{cite}
\usepackage{hyperref}
\hypersetup{hidelinks}

\newcommand{\tabcite}[1]{\textsuperscript{\cite{#1}}}
\newcommand{\fullmark}{\ensuremath{\checkmark}}   
\newcommand{\partmark}{\ensuremath{\bigcirc}}     
\newcommand{\nomark}{\ensuremath{\times}}         

\newlength{\panelwd}
\newlength{\panelht}
\newcommand{\panellabelin}[2]{%
  \setlength{\panelwd}{\linewidth}%
  \begin{tikzpicture}
    \node[anchor=south west,inner sep=0] (img)
         {\includegraphics[width=\panelwd]{#1}};
    \node[anchor=north west, inner sep=2.5pt, font=\figfont\normalsize\bfseries, text=black]
         at (img.north west) {#2};
  \end{tikzpicture}%
}
\newcommand{\panellabel}[2]{%
  \setlength{\panelwd}{\linewidth}%
  \begin{tikzpicture}
    \node[anchor=south west,inner sep=0] (img)
         {\includegraphics[width=\panelwd]{#1}};
    \node[overlay, anchor=north west, inner sep=1pt, font=\figfont\normalsize\bfseries, text=black]
         at ([shift={(-1pt,1pt)}]img.north west) {#2};
  \end{tikzpicture}%
}
\newcommand{\panellabelside}[2]{%
  \setlength{\panelht}{0.52\linewidth}%
  \begin{tikzpicture}
    \node[anchor=south,inner sep=0] (img) at (0.57\linewidth,0)
         {\includegraphics[height=\panelht]{#1}};
    \node[anchor=north west, inner sep=0pt, font=\figfont\normalsize\bfseries, text=black]
         at (0,0 |- img.north) {#2};
    \path[use as bounding box] (0,0) rectangle (\linewidth,0 |- img.north);
  \end{tikzpicture}%
}

\title{Fully integrated continuous-variable quantum key distribution with composable security over 100 km}

\author[1,†]{Yankai Xu}
\author[1,†]{Xinhang Li}
\author[1,2,3,*]{Tao Wang}
\author[5]{Jisheng Dai}
\author[5]{Xueqin Jiang}
\author[1]{Yuyao Guo}
\author[1,2,3]{Peng Huang}
\author[1,4]{Linjie Zhou}
\author[1,2,3,*]{Guihua Zeng}

\affil[1]{State Key Laboratory of Photonics and Communications, Shanghai Jiao Tong University, Shanghai 200240, China}
\affil[2]{Shanghai Research Center for Quantum Sciences, Shanghai 201315, China}
\affil[3]{Hefei National Laboratory, Hefei 230088, China}
\affil[4]{SJTU-Pinghu Institute of Intelligent Optoelectronics, Pinghu 314200, China}
\affil[5]{College of Information Science and Technology, Donghua University, Shanghai 201315, China}

\affil[*]{Corresponding authors: tonystar@sjtu.edu.cn, ghzeng@sjtu.edu.cn}
\affil[†]{These authors contributed equally to this work.}
\date{}

\begin{document}

\maketitle

\begin{abstract}
\noindent
Quantum key distribution (QKD) guarantees information-theoretic security by the laws of physics, but deployment at scale requires compact, manufacturable photonic terminals. Continuous-variable QKD (CV-QKD) is well suited for this transition through telecom-compatible, room-temperature coherent detection. However, unifying full on-chip core terminal integration, room-temperature operation, high loss tolerance, and composable end-to-end security in long-distance QKD remains a key bottleneck. Here we report a fully integrated CV-QKD platform in which two hybrid III--V/Si$_3$N$_4$ integrated lasers, a silicon transmitter, and a silicon coherent receiver implement the core terminal functions, operating with a local local oscillator (LLO) over fibre links of 25--150~km. A Bayesian machine-learning algorithm maintains robust phase lock throughout the long records required for composable security, consistently outperforming the conventional unscented Kalman filter, while rate-matched multidimensional reconciliation approaches the Shannon limit. The system certifies a composable finite-size secret-key rate of 29.3~kbps at 100~km from a 140-billion-symbol block, with 12.9~kbps at 125 km under finite-size analysis and 9.17~kbps at 150 km under asymptotic analysis. By establishing the longest finite-size and asymptotic reaches and the highest secret-key rate per symbol reported for integrated CV-QKD, this work advances the development of practical chip-based quantum networks.

\end{abstract}

\section{Introduction}

Quantum key distribution (QKD) delivers secret keys whose security does not depend on assumptions about an adversary's computing power~\cite{ref1,ref2}, and it has already progressed from single links into metropolitan and high-rate networks~\cite{ref6,ref7,ref11,ref12}.  Scaling QKD to the network level now hinges on the terminal, whose user-end hardware must shrink from bulky, costly optical devices to a manufacturable chip, while links extend from metropolitan towards inter-city scales.

Towards this end, chip-based QKD has advanced furthest in discrete-variable (DV) systems, including integrated transmitters and decoders~\cite{Ma2016,Sibson2017a,Sibson2017b}, detector-integrated receiver chips~\cite{Beutel2021,Beutel2022}, standalone photonic integrated circuit (PIC)-based secure communication~\cite{Paraiso2021}, and high-speed long-distance links~\cite{Dolphin2023,Sax2023,Zhang2026FiveGHz}. Their reach, however, remains constrained by single-photon detectors (SPDs). The highest-efficiency devices are superconducting nanowire SPDs (SNSPDs) that require cryogenic cooling~\cite{Li2025SNSPD99}, whereas room-temperature avalanche photodiodes (APDs) trade efficiency for higher dark counts and afterpulsing~\cite{Sax2023}.

CV-QKD avoids this detector bottleneck, because it encodes the key in the quadratures of coherent states~\cite{ref15,ref14,Weedbrook2012Gaussian} and measures them with room-temperature coherent detection already used in classical optical networks~\cite{chip5,ref4,ref5}. A complete CV-QKD terminal requires lasers, an in-phase/quadrature (IQ) modulator, and a balanced coherent receiver, all of which can in principle be integrated on chip~\cite{chip2,chip5}. Compared to a transmitted local oscillator (TLO) alongside the quantum signal, operating with a local local oscillator (LLO) removes a known side channel and matches standard coherent-receiver practice~\cite{exp3,Soh2015SelfRefCVQKD,ref19}. Integrated transmitters, receivers, high-symbol-rate links, and on-chip sources have been demonstrated separately~\cite{chip2,chip3,chip5,chip6,chip7,chip8,chip9,chip10,chip13}, and 100~km-class LLO operation has so far relied on bulk transmitters and receivers~\cite{ref19,PP2024ImprovedComposable}. While composable finite-size security has been demonstrated in chip-based CV-QKD, its reach remains in short transmission distances~\cite{ref17,ref18}. Taken together, these advances leave it an open challenge whether a fully integrated system comprising lasers, a transmitter, and a coherent receiver can operate over a long fibre link while simultaneously executing parameter estimation, reconciliation, and composable security analysis.

Fundamentally, photonic integration and long-distance transmission strain the same excess-noise margin, which diminishes as channel loss grows~\cite{ref21,MeleMaxExcessNoise2024}. Integrated lasers, modulators, and detectors, which are typically less stable than their discrete bulk-optics counterparts, consume this budget through wavelength drift, power drift, residual preparation error, and detector imperfections~\cite{chip8}. An ultra-narrow-linewidth integrated laser suppressing the free-running phase diffusion, stabilised biasing transmitter, and calibrated balanced detection reduce one part of the cost, but they do not remove the need for pilot-aided recovery. Crucially, in an LLO configuration, the receiver must estimate the relative phase between two free-running chip lasers. Over a long fiber link, the accuracy of this estimation determines whether any secret key can be successfully generated. The carrier-phase recovery consumes the budget further, because the pilot that carries the phase is weak and shares the digitiser with the quantum signal.  At the low signal-to-noise ratio (SNR) of a 100~km link, a slight pilot error can break phase lock before key extraction begins~\cite{ref41}. Furthermore, this lock must also survive the long records required for finite-size composable security~\cite{ref47}, while reconciliation operates close to channel capacity. Extracting a valid secret key therefore requires combining device-level noise suppression with algorithmic recovery to ensure composable security across a continuous data record.

Here, we demonstrate the first full-link implementation of a fully integrated CV-QKD system containing all core functions, in which the integrated lasers, transmitter, and receiver are co-designed with the complete digital chain from phase recovery through reconciliation to privacy amplification. Ten-Hz-level integrated lasers keep relative phase diffusion within the regime trackable by a phase-frequency state-space model, so pilot observations across the record can jointly constrain the recovered phase trajectory. This machine-learning-based algorithm maintains robust lock throughout the long records required for composable security while achieving a lower phase-estimation mean-square error than the standard unscented Kalman filter (UKF). Rate-matched multidimensional reconciliation operates near the Shannon limit and interleaved quantum and vacuum acquisitions maintain the shot-noise reference under identical receiver settings. The reconciliation stage assigns each codeword to a rate-matched decoding route based on its publicly disclosed SNR. At 100~km, composable finite-size analysis against collective attacks certifies a secret-key rate of 29.3~kbps from a measured block of 140 billion symbols~\cite{ref17,ref18,PP2024ImprovedComposable}. The system also achieves secret-key rates of 12.9~kbps at 125~km under finite-size analysis and 9.17~kbps at 150~km under asymptotic analysis. Collectively, these results establish the longest reported reaches for integrated CV-QKD in their respective security regimes. On a per-channel-use basis, this is the highest reported secret-key rate among chip-based CV- and DV-QKD demonstrations at this distance, successfully translating full photonic integration into practical, long-distance secret-key generation.

\begin{table}[!bp]
  \centering
  \caption{\textbf{Single-link chip-QKD demonstrations.}}
  \label{tab:sota}
  \begin{threeparttable}
    \footnotesize
    \setlength{\tabcolsep}{0.35pt}
    \renewcommand{\arraystretch}{1.1}
    \begin{tabular}{@{}ccccccccccccc@{}}
      \toprule
      \multicolumn{1}{c}{\textbf{Ref.}} & \multicolumn{1}{c}{\textbf{Sec.}} & \multicolumn{1}{c}{\textbf{\boldmath$R_s$}} & \multicolumn{1}{c}{\textbf{Mod.}} & \multicolumn{1}{c}{\textbf{Integration }} & \multicolumn{1}{c}{\textbf{\boldmath$L$}} & \multicolumn{1}{c}{\textbf{Loss}} & \multicolumn{1}{c}{\textbf{\boldmath$K_\infty$}} & \multicolumn{1}{c}{\textbf{\boldmath$K_{\mathrm{fin}}$}} & \multicolumn{1}{c}{\textbf{\boldmath$N$}} & \multicolumn{1}{c}{\hspace{1.5pt}\textbf{CS}\hspace{1.5pt}} & \multicolumn{1}{c}{\hspace{1.5pt}\textbf{PP}\hspace{1.5pt}} & \multicolumn{1}{c}{\hspace{1pt}\textbf{Full}} \\
      \midrule
      \multirow{6}{*}{\shortstack[c]{\textbf{This}\\\textbf{work}}}
        & \multirow{6}{*}{Finite} & \multirow{6}{*}{0.1} & \multirow{6}{*}{Gaussian} & \multirow{6}{*}{\shortstack[c]{InP/Si$_3$N$_4$ lasers\\{}+ Si Tx + Si Rx}}
                                                          & 25.25  & 4.1  & $7.86\!\times\!10^{-2}$ & $6.81\!\times\!10^{-2}$ & $10^{8}$               & \multirow{6}{*}{\fullmark} & \multirow{6}{*}{\fullmark} & \multirow{6}{*}{\fullmark} \\
        & & & & & 50.4  & 8.1  & $2.13\!\times\!10^{-2}$ & $1.26\!\times\!10^{-2}$ & $10^{8}$               & & & \\
        & & & & & 75.65  & 12.2 & $7.85\!\times\!10^{-3}$ & $5.01\!\times\!10^{-3}$ & $10^{9}$               & & & \\
        & & & & & 100.8 & 16.2 & $1.11\!\times\!10^{-3}$ & $2.93\!\times\!10^{-4}$ & $1.4\!\times\!10^{11}$ & & & \\
        & & & & & 126.06 & 20.3 & $4.70\!\times\!10^{-4}$ & $1.29\!\times\!10^{-4}$ & $5.9\!\times\!10^{10}$ & & & \\
        & & & & & 151.32 & 24.4 & $9.17\!\times\!10^{-5}$ & --                      & $1.5\!\times\!10^{8}$  & & & \\
      \cmidrule(l){1-13}
      \tabcite{Xu2026Bidir}
        & Finite & 1 & Gaussian/QPSK & Si Tx + Si Rx & 65.5 & 11.8 & -- & $1.41\!\times\!10^{-3}$ & $2\!\times\!10^{9}$ & \nomark & \nomark & \nomark \\
      \addlinespace[1.5pt]
      \tabcite{chip10}
        & Finite & 40 & DP-QPSK & Si Tx + Si Rx & 10 & 1.98 & -- & $3.03\!\times\!10^{-2}$ & $\sim\!10^{10}$ & \fullmark & \fullmark & \nomark \\
      \addlinespace[1.5pt]
      \multirow{2}{*}{\tabcite{chip3}}
        & \multirow{2}{*}{Finite} & \multirow{2}{*}{16} & \multirow{2}{*}{PCS-QAM} & \multirow{2}{*}{Si Tx + Si Rx}
                                                          & 10 & 2.0 & -- & $3.34\!\times\!10^{-2}$ & $2\!\times\!10^{6}$ & \nomark & \nomark & \nomark \\
        & & & & & 20 & 4.0 & -- & $1.54\!\times\!10^{-2}$ & $2\!\times\!10^{7}$ & \nomark & \nomark & \nomark \\
      \addlinespace[1.5pt]
      \multirow{2}{*}{\tabcite{chip5}}
        & \multirow{2}{*}{Finite} & \multirow{2}{*}{10} & \multirow{2}{*}{PCS-QAM} & \multirow{2}{*}{Si Rx}
                                                          & 5  & 1.0 & -- & $9.3\!\times\!10^{-2}$  & $1.6\!\times\!10^{7}$ & \nomark & \nomark & \nomark \\
        & & & & & 10 & 2.0 & -- & $3.51\!\times\!10^{-2}$ & $1.6\!\times\!10^{7}$ & \nomark & \nomark & \nomark \\
      \addlinespace[1.5pt]
      \tabcite{chip7}
        & Asymp. & 1.56 & 8-PSK & Si Tx + Si Rx & 50.4 & 10.1 & $3.23\!\times\!10^{-3}$ & -- & -- & \nomark & \nomark & \nomark \\
      \addlinespace[1.5pt]
      \tabcite{chip9}
        & Asymp. & 0.1 & Gaussian & Si Rx & 23 & 4.6 & $2.20\!\times\!10^{-3}$ & -- & -- & \nomark & \nomark & \nomark \\
      \addlinespace[1.5pt]
      \tabcite{chip6}
        & Asymp. & 1 & Gaussian & Si Rx & 28.6 & 5.7 & $1.38\!\times\!10^{-3}$ & -- & -- & \nomark & \nomark & \nomark \\
      \addlinespace[1.5pt]
      \tabcite{chip13}
        & Asymp. & 0.016 & Gaussian & InP Tx & 11 & 2.04 & $4.88\!\times\!10^{-3}$ & -- & -- & \nomark & \nomark & \nomark \\
      \addlinespace[1.5pt]
      \tabcite{chip8}
        & Asymp. & 0.25 & Gaussian & InP/Si$_3$N$_4$ lasers & 50 & -- & $3.00\!\times\!10^{-3}$ & -- & -- & \nomark & \nomark & \nomark \\
      \addlinespace[1.5pt]
      \tabcite{chip2}
        & Asymp. & $0.0008$ & Gaussian & Si Tx + Si Rx & 100 & 16.0 & $1.80\!\times\!10^{-4}$ & -- & -- & \nomark & \nomark & \nomark \\
      \cmidrule(l){1-13}
      \multirow{2}{*}{\tabcite{ref6}}
        & \multirow{2}{*}{Finite} & \multirow{2}{*}{2.5} & \multirow{2}{*}{Decoy BB84 pol.} & \multirow{2}{*}{Si Tx}
                                                          & 10  & 2.2  & -- & $4.63\!\times\!10^{-2}$          & $10^{8}$ & \multirow{2}{*}{\fullmark} & \multirow{2}{*}{\fullmark} & \multirow{2}{*}{\nomark} \\
        & & & & & 328 & 55.1 & -- & $\!9.3\!\times\!10^{-8}$ & $10^{8}$ & & & \\
      \addlinespace[1.5pt]
      \tabcite{Sibson2017b}
        & Asymp. & 0.56 & Decoy BB84 time-bin & InP Tx + SiON Rx & 20 & 4.0 & $6.16\!\times\!10^{-4}$ & -- & -- & \nomark & \nomark & \partmark \\
      \addlinespace[1.5pt]
      \tabcite{Paraiso2021}
        & Asymp. & 1 & Decoy BB84 phase & InP Tx + SiON Rx & 50 & 10.0 & $2.80\!\times\!10^{-5}$ & -- & -- & \nomark & \fullmark & \partmark \\
      \addlinespace[1.5pt]
      \tabcite{Dolphin2023}
        & Finite & 1 & Decoy BB84 phase & InP/SiN transceiver & 250 & 44.0 & $1.86\!\times\!10^{-7}$ & $6.70\!\times\!10^{-8}$ & $1.04\!\times\!10^{7}$ & \nomark & \nomark & \nomark \\
      \addlinespace[1.5pt]
      \tabcite{Sax2023}
        & Finite & 2.5 & Decoy BB84 time-bin & Si Tx + FLW-glass Rx & 202.0 & 39.5 & -- & $3.76\!\times\!10^{-6}$ & -- & \fullmark & \fullmark & \nomark \\
      \addlinespace[1.5pt]
      \tabcite{Wei2023}
        & Finite & 0.05 & Decoy BB84 pol. & Si Tx + Si Rx & 150 & 29.0 & -- & $1.73\!\times\!10^{-5}$ & $\sim\!10^{7}$ & \fullmark & \nomark & \nomark \\
      \addlinespace[1.5pt]
      \tabcite{Zhang2026FiveGHz}
        & Finite & 5 & Decoy BB84 pol. & Si Tx + SNSPD chip & 200 & 36.0 & -- & $2.10\!\times\!10^{-5}$ & -- & \fullmark & \nomark & \nomark \\
      \tabcite{Semenenko2020MDI}
        & Asymp. & 0.25 & MDI decoy BB84 & InP Tx + InP Tx & -- & 20.0 & $4.00\!\times\!10^{-6}$ & -- & -- & \nomark & \nomark & \nomark \\
      \addlinespace[1.5pt]
      \tabcite{Du2024TF}
        & Finite & 0.5 & TWCC-SNS TF & InP Tx + InP Tx & -- & 50.0 & -- & $2\!\times\!10^{-6}$ & $10^{8}$ & \nomark & \nomark & \nomark \\
      \bottomrule
    \end{tabular}
    \begin{tablenotes}
    \scriptsize
    \item[] Sec., security analysis regime; $R_s$, symbol rate in GBaud; $L$, fibre distance in km; Loss, channel loss in dB; $K_\infty$ and $K_{\mathrm{fin}}$, asymptotic and finite-size key rates in bits/symbol; $N$, block size. CS and PP denote composable security and post-processing, respectively. For integration status: \fullmark{} (Full) indicates simultaneous on-chip local laser sources, Tx, and Rx in a single end-to-end LLO link demonstration; \partmark{} (Partial), all present with off-chip subfunctions; and \nomark{}, at least one core component absent.
    \end{tablenotes}
  \end{threeparttable}
\end{table}

\section{Results}\label{sec:results}

\subsection{Integrated transceiver system}\label{sec:platform}
\begin{figure}[!ht]
\centering
\begin{subfigure}[b]{0.72\linewidth}
    \centering
    \panellabelin{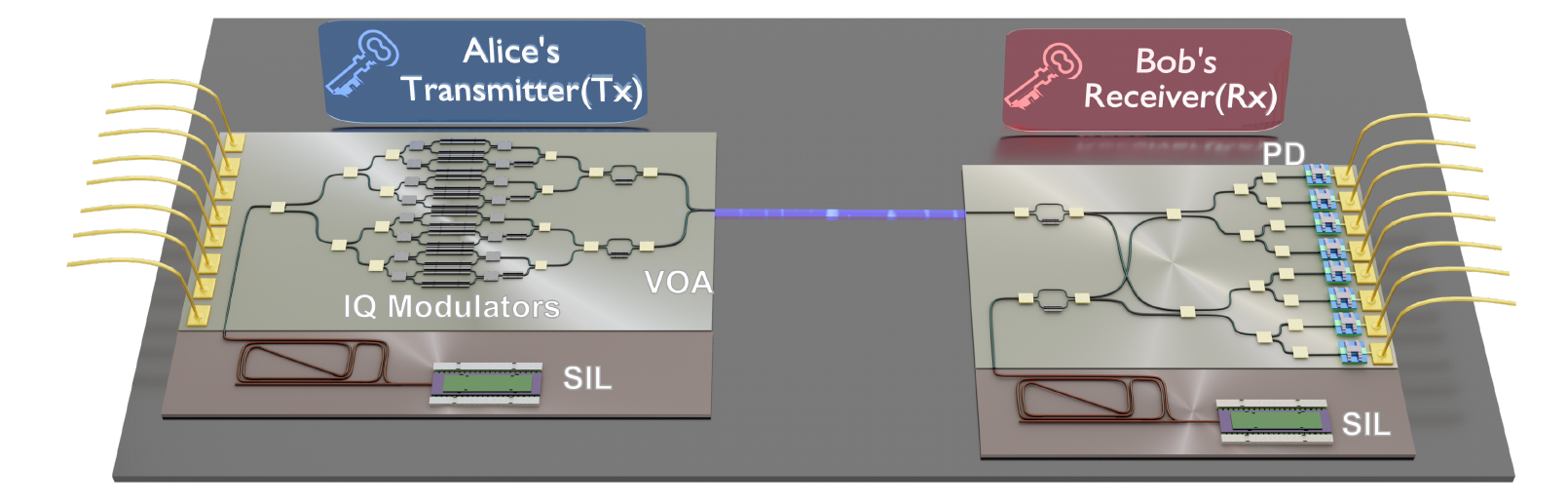}{a}
\end{subfigure}\\[0pt]
\begin{subfigure}[b]{0.329\linewidth}
    \centering
    \panellabelside{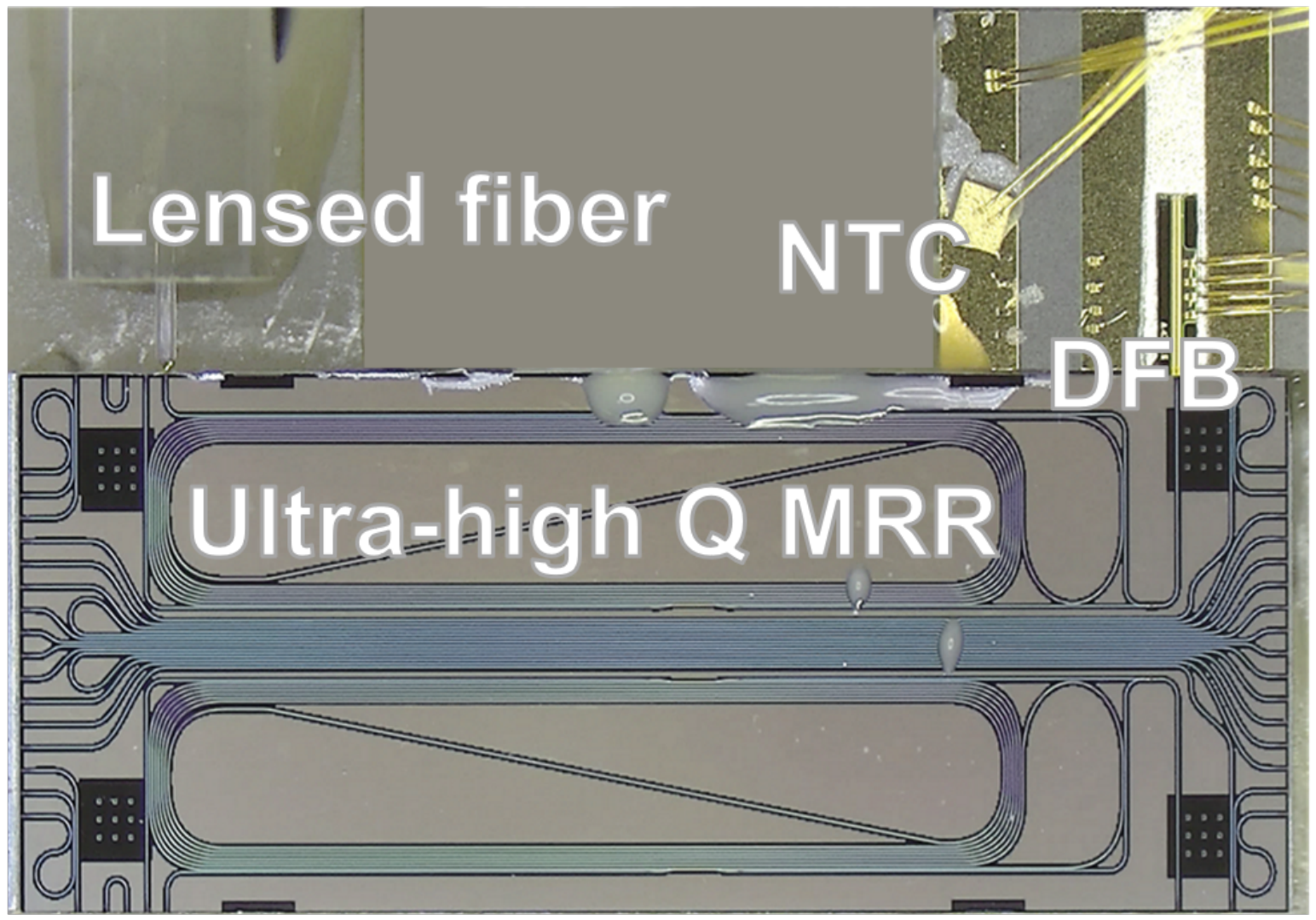}{b}
\end{subfigure}\hspace{0.003\linewidth}%
\begin{subfigure}[b]{0.329\linewidth}
    \centering
    \panellabelside{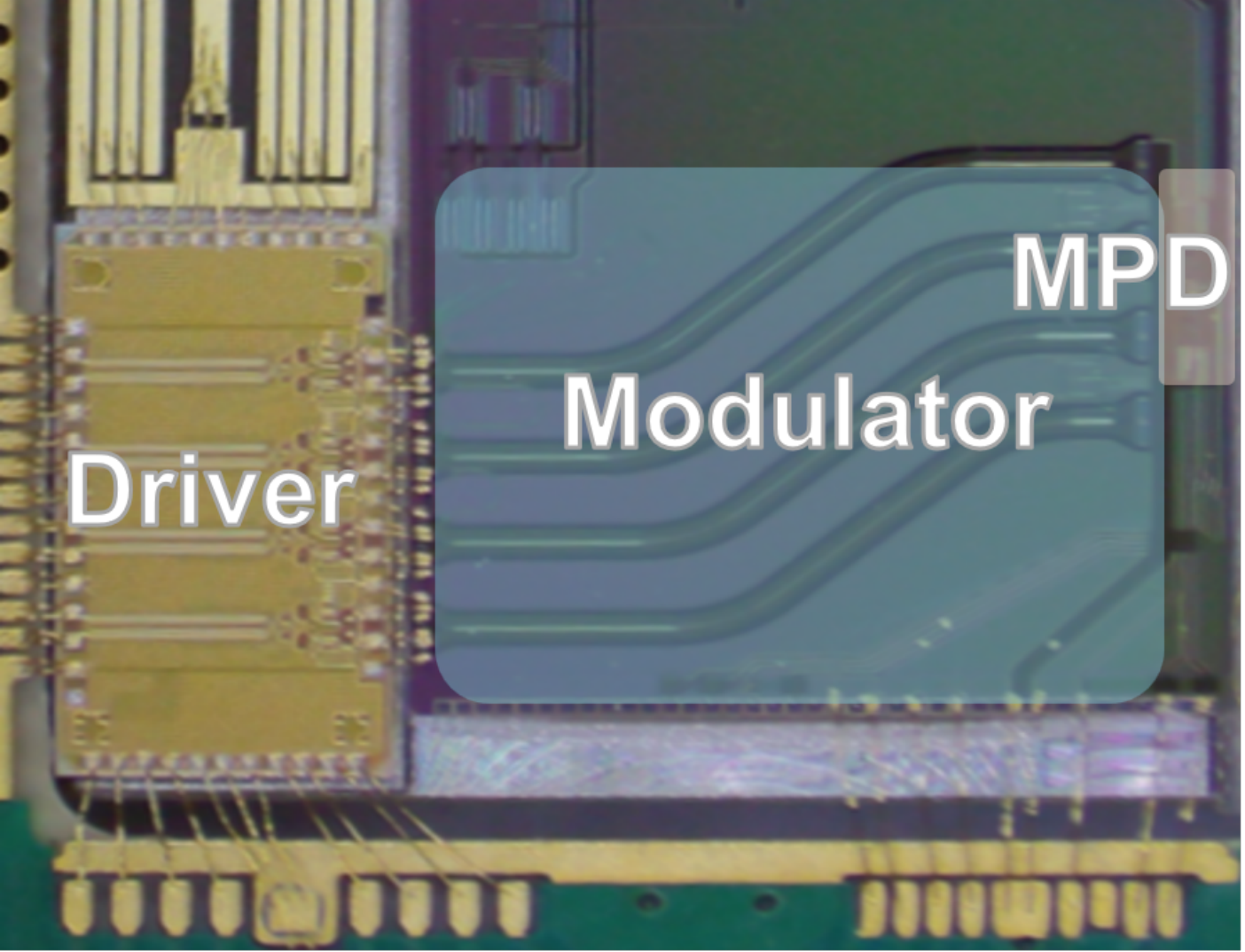}{c}
\end{subfigure}\hspace{0.003\linewidth}%
\begin{subfigure}[b]{0.329\linewidth}
    \centering
    \panellabelside{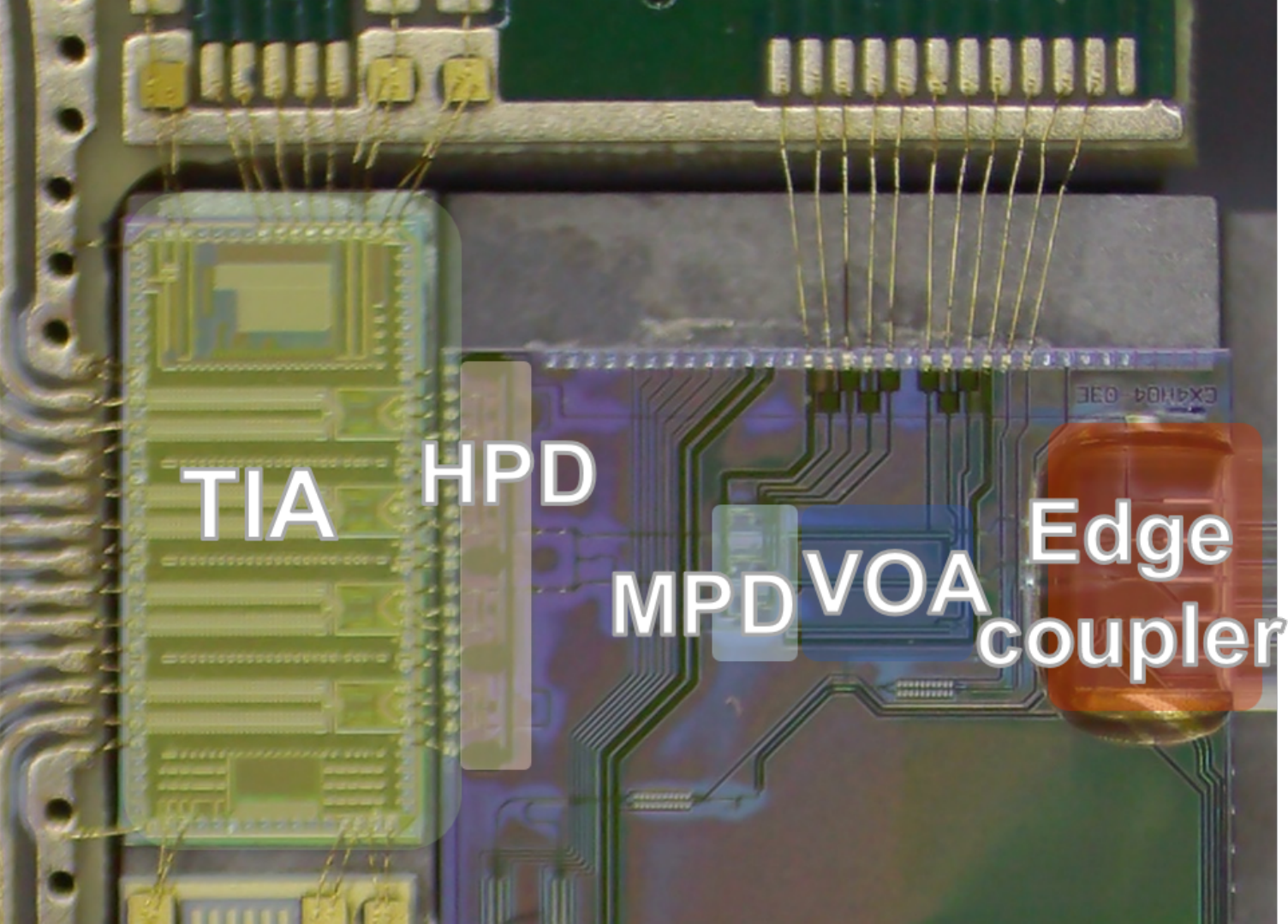}{d}
\end{subfigure}\\[0pt]
\begin{subfigure}[b]{0.329\linewidth}
    \centering
    \panellabel{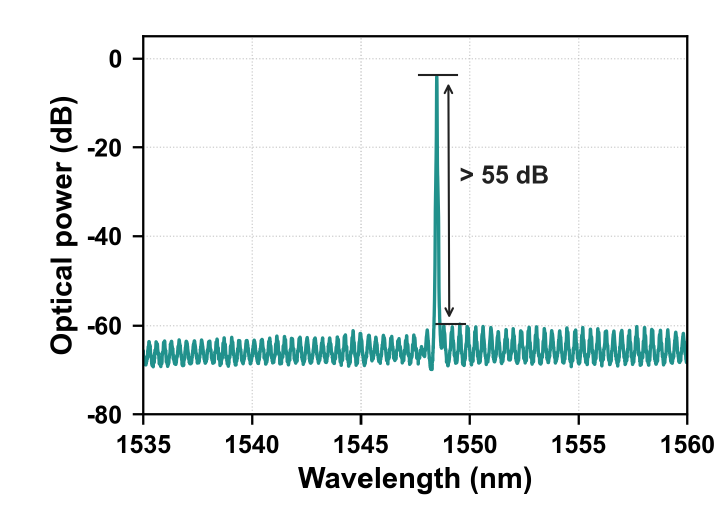}{e}
\end{subfigure}\hspace{0.003\linewidth}%
\begin{subfigure}[b]{0.329\linewidth}
    \centering
    \panellabel{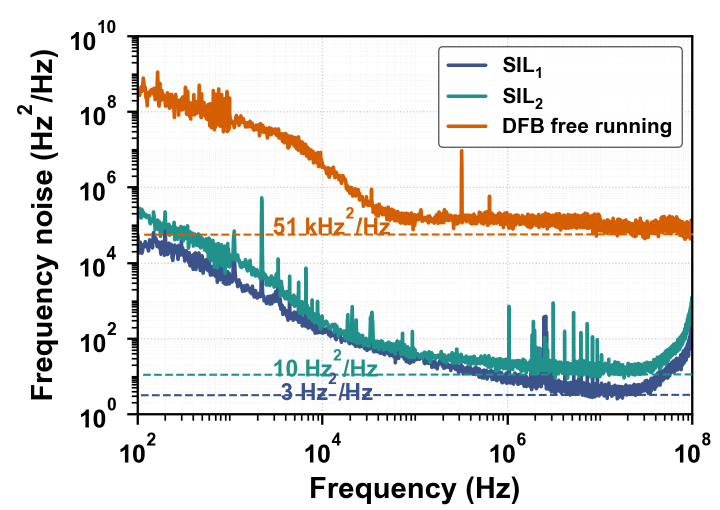}{f}
\end{subfigure}\hspace{0.003\linewidth}%
\begin{subfigure}[b]{0.329\linewidth}
    \centering
    \panellabel{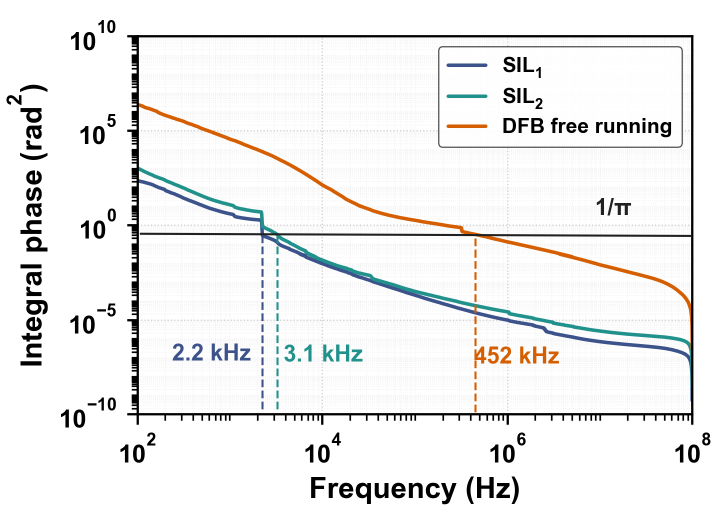}{g}
\end{subfigure}\\[0pt]
\begin{subfigure}[b]{0.325\linewidth}
    \centering
    \panellabel{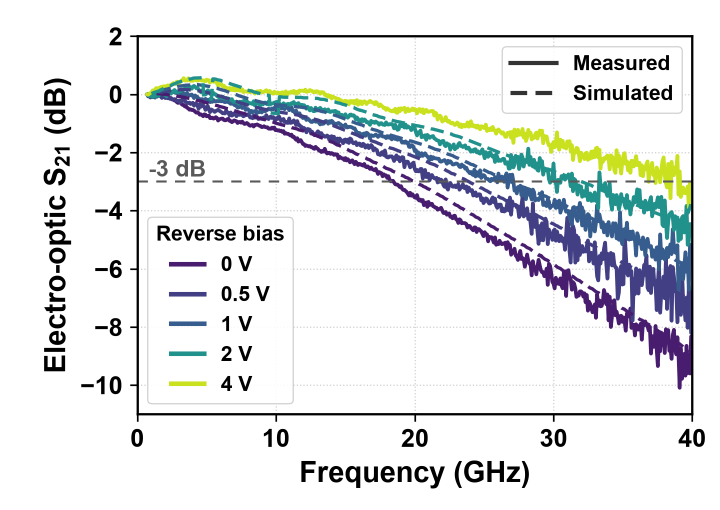}{h}
\end{subfigure}\hspace{0.008\linewidth}%
\begin{subfigure}[b]{0.325\linewidth}
    \centering
    \panellabel{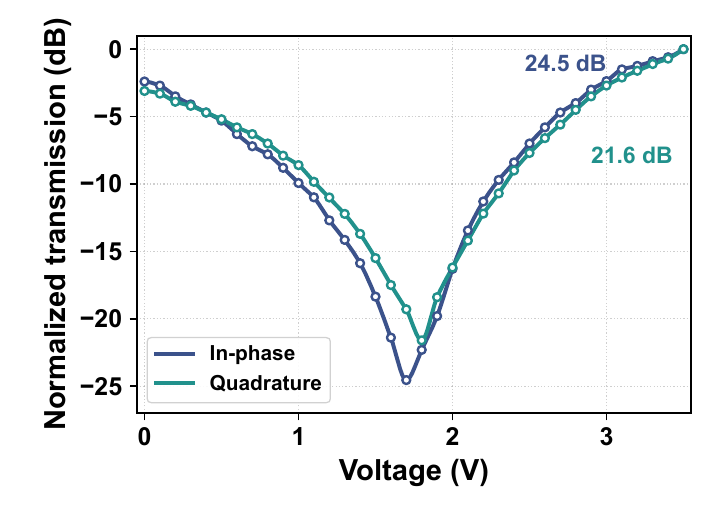}{i}
\end{subfigure}\hspace{0.008\linewidth}%
\begin{subfigure}[b]{0.325\linewidth}
    \centering
    \panellabel{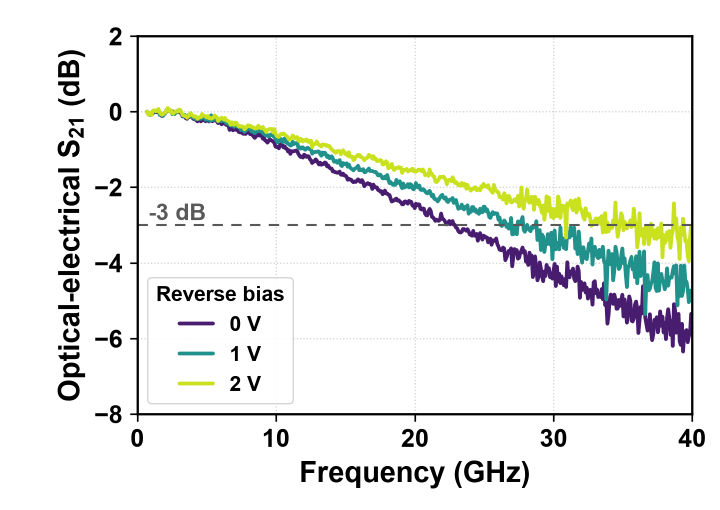}{j}
\end{subfigure}
\par\vspace{2pt}
\caption[Integrated platform]{\textbf{Characteristics of integrated transceiver system.}  (a) System schematic of Alice's transmitter and Bob's receiver connected by the quantum link.  (b) Hybrid SIL laser package, showing lensed-fibre coupling, the negative temperature coefficient (NTC) thermistor, the DFB gain chip, and the ultra-high-$Q$ MRR feedback chip.  (c) Transmitter package with the quad-channel driver, silicon modulator, and monitor photodiode.  (d) Receiver package with the TIA, hybrid photodiode, monitor photodiode, and variable optical attenuator.  (e) Optical spectrum of the SIL laser, showing an SMSR exceeding 55~dB.  (f) Frequency-noise spectra of the two SIL lasers and the free-running DFB reference.  (g) Integral phase noise, giving integral linewidths of 2.2 and 3.1~kHz for the SIL lasers and 452~kHz for the free-running DFB.  (h) Electro-optic $S_{21}$ of the silicon modulator at 0 to 4~V reverse bias from bottom to top, measured (solid) and modelled as a terminated transmission line over the 0 to 2~V range (dashed, Supplementary Note~5.1).  (i) Voltage-dependent transmission of the in-phase and quadrature MZI.  (j) Optical-electrical $S_{21}$ of the photodiode at 0, 1, and 2~V reverse bias from bottom to top (Supplementary Note~5.1).}
\label{fig:chip-platform}
\end{figure}

Fig.~\ref{fig:chip-platform} illustrates the integrated LLO CV-QKD platform and the characterization of its principal photonic components. The link comprises a transmitter node (Alice) and a receiver node (Bob), with the core terminal functions implemented by two integrated lasers, a silicon transmitter and a silicon coherent receiver.

The optical carrier and the local oscillator are generated by two hybrid III--V/Si$_3$N$_4$ self-injection-locking (SIL) lasers designed for the ultra-low noise required in long-distance CV-QKD. Each laser butt-couples a commercial distributed feedback (DFB) gain chip to a low-loss silicon-nitride feedback circuit, in which a high-$Q$ microring resonator (MRR) is inserted into a Sagnac loop. The MRR employs an add-drop configuration with a 3-$\mu$m-wide waveguide. The expanded mode field reduces the optical intensity at the interfaces and hence the propagation loss, while Euler bends smooth the transition between straight and curved sections to suppress higher-order modes within a compact footprint. With this high-$Q$ feedback chip (loaded $Q$ of $1.4\times10^{6}$, design details in Supplementary Note~5.2), the integrated laser achieves a side-mode suppression ratio (SMSR) exceeding 55~dB in the SIL state, as shown in Fig.~\ref{fig:chip-platform}(e). The frequency-noise spectra in Fig.~\ref{fig:chip-platform}(f) give intrinsic linewidths of 10 and 30~Hz for the two SIL lasers, compressed by more than 5,000-fold from the 160~kHz free-running DFB linewidth. The integral linewidths that govern the long-term phase stability of the link are 2.2 and 3.1~kHz, as shown in Fig.~\ref{fig:chip-platform}(g). One laser at Alice supplies the coherent-state carrier and the other runs free at Bob as the local oscillator, so the receiver operates with an LLO and no bright phase reference is sent through the fibre. This narrow linewidth ensures that the relative phase noise of the two lasers remains within the tracking range of the public-pilot phase recovery described in Section~\ref{sec:phase-recovery}.

The transmitter and receiver each employ a system-in-package (SiP) platform in which a silicon-on-insulator (SOI) photonic integrated circuit (PIC) is co-packaged with essential high-speed electronics. These electronics include a quad-channel modulator driver dedicated to state preparation at Alice and a linear transimpedance amplifier (TIA) for coherent detection at Bob. Each integrated module is mounted as a ball-grid-array (BGA) assembly onto a custom control board (detailed in Supplementary Note~5). Ultimately, the packaged transmitter and receiver modules exhibit 3-dB modulation and detection bandwidths above 30 and 35~GHz, respectively, far beyond the 100~MBaud symbol rate, so their frequency responses are flat across the quantum band. The electro-optic response of the modulator and optical-electrical response of the photodiode (PD) are plotted in Fig.~\ref{fig:chip-platform}(h,j), and the voltage-dependent transmission of the in-phase and quadrature Mach--Zehnder interferometers is shown in Fig.~\ref{fig:chip-platform}(i), with measured extinction ratios of 24.5 and 21.6~dB respectively. The two arms hold a phase error within $5^\circ$ and time skew within 3~ps, which strictly bounds the transmitter-preparation contribution to the excess-noise budget detailed in Section~\ref{sec:excess-noise}. On the detection side, the balanced receiver provides a common-mode rejection of at least 20 dB at low frequency and 16 dB across the radio-frequency band, with an overall detection efficiency of 40.8\%.

\subsection{Noise analysis}\label{sec:excess-noise}

To identify the dominant noise source in the integrated platform, we grouped the excess-noise contributions, including relative intensity noise (RIN) and residual phase noise (RPN), as

\begin{equation}
\xi=\xi_{\mathrm{RIN}}+\xi_{\mathrm{mod}}
+\xi_{\mathrm{RPN}}
+\xi_{\mathrm{other}} .
\label{eq:excess-sum}
\end{equation}

Here $\xi_{\mathrm{RIN}}$ represents the laser RIN contribution, $\xi_{\mathrm{mod}}$ is the transmitter preparation noise, and $\xi_{\mathrm{RPN}}$ denotes the residual phase noise.  Digital-to-analogue converter (DAC) and electrical-drive noise are included in $\xi_{\mathrm{mod}}$. The residual term \(\xi_{\mathrm{other}}\) accounts for additive leakage, receiver digitisation noise, and calibration residuals, which are minor.  Supplementary Note~3 defines the modelled terms in detail.

At 100~km, the measured signal-laser RIN of $-150$~dBc/Hz over $B_{\mathrm q}\simeq100$~MHz yields $\xi_{\mathrm{RIN,sig}}=1.7\times10^{-7}$~shot-noise units (SNU) at $V_A=6.77$.  The packaged-module drive SNR of approximately 55~dB contributes $\xi_{\mathrm{mod}}\simeq2.1\times10^{-5}$~SNU. A model-based diagnostic gave \(V_{\mathrm{est}}=1.542\times10^{-3}\)~rad\(^{2}\). Accounting for intrinsic laser-linewidth diffusion increased the total residual phase variance slightly to \(\sigma_\phi^2=1.545\times10^{-3}\)~rad\(^{2}\). This variance corresponds to a residual-phase-noise contribution of \(\xi_{\mathrm{RPN}}=1.046\times10^{-2}\)~SNU, the largest term in the modelled excess-noise budget. Because \(\xi_{\mathrm{RPN}}\) scales approximately with \(V_A\), reducing the modulation variance lowers the phase noise but also decreases the received SNR and increases reconciliation failures. Based on a parameter sweep constrained by the pilot SNR, detector noise, and available code rates, we selected \(V_A=6.77\) for the 100-km link.

With this operating point fixed, we propagated the measured phase-MSE advantage of URTS through the 100~km excess-noise model to assess its impact on composable security. Using the masked-pilot comparison described in Section~\ref{sec:phase-recovery}, URTS achieved a lower MSE in all randomly sampled acquisitions, giving a UKF-to-URTS MSE ratio of 1.1097 (95\% CI, 1.1018--1.1191; 2,000 frames). This mean ratio corresponds to a UKF-equivalent excess noise of \(\bar{\xi}_{\mathrm{UKF}}=1.196\times10^{-2}\)~SNU, compared with \(\bar{\xi}_{\mathrm{URTS}}=1.078\times10^{-2}\)~SNU. At the RPN level, we used the largest observed MSE reduction of \(32.47\%\) for the worst-case analysis. Starting from \(V_{\mathrm{est}}^{\mathrm{URTS}}=1.542\times10^{-3}\)~rad\(^{2}\) and retaining \(V_{\mathrm{drift}}=2.51\times10^{-6}\)~rad\(^{2}\) gives \(V_{\mathrm{est}}^{\mathrm{UKF}}=2.283\times10^{-3}\)~rad\(^{2}\). This corresponds to \(\xi_{\mathrm{RPN}}^{\mathrm{UKF}}=1.549\times10^{-2}\)~SNU. The resulting excess noise upper bound is \(\xi_{\mathrm{PE(UKF)}}^U=2.443\times10^{-2}\)~SNU. Substituting this inferred bound into the otherwise unchanged finite-size analysis yields no positive composable lower bound on the secret-key rate. Consequently, the advantage ultimately preserves the critical excess-noise margin required to extract a positive finite-size composable key at 100~km.

\begin{figure}[!htbp]
\centering
\includegraphics[width=0.9\linewidth]{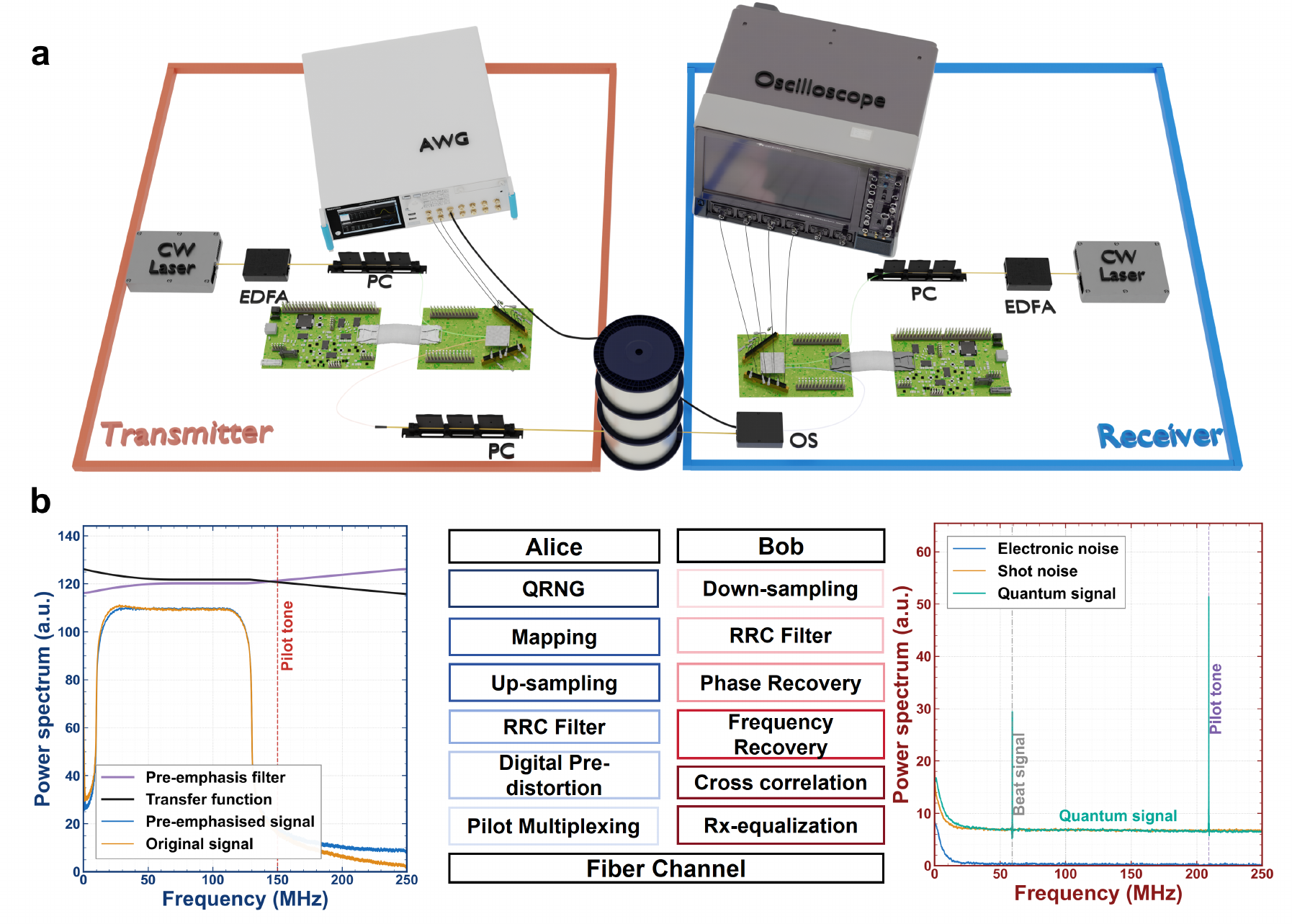}
\caption{\textbf{Integrated LLO CV-QKD system and DSP chain.} (a) Alice's transmitter and Bob's receiver, including the integrated laser modules, silicon transmitter, silicon receiver, erbium-doped fibre amplifier (EDFA), polarization controllers (PCs), optical switch (OS), and fibre spools. (b) Transmitter and receiver DSP chains.}
\label{fig:system-dsp}
\end{figure}

\subsection{Experimental investigation}\label{sec:setup}
\begin{figure}[!htbp]
\centering
\includegraphics[width=0.86\linewidth]{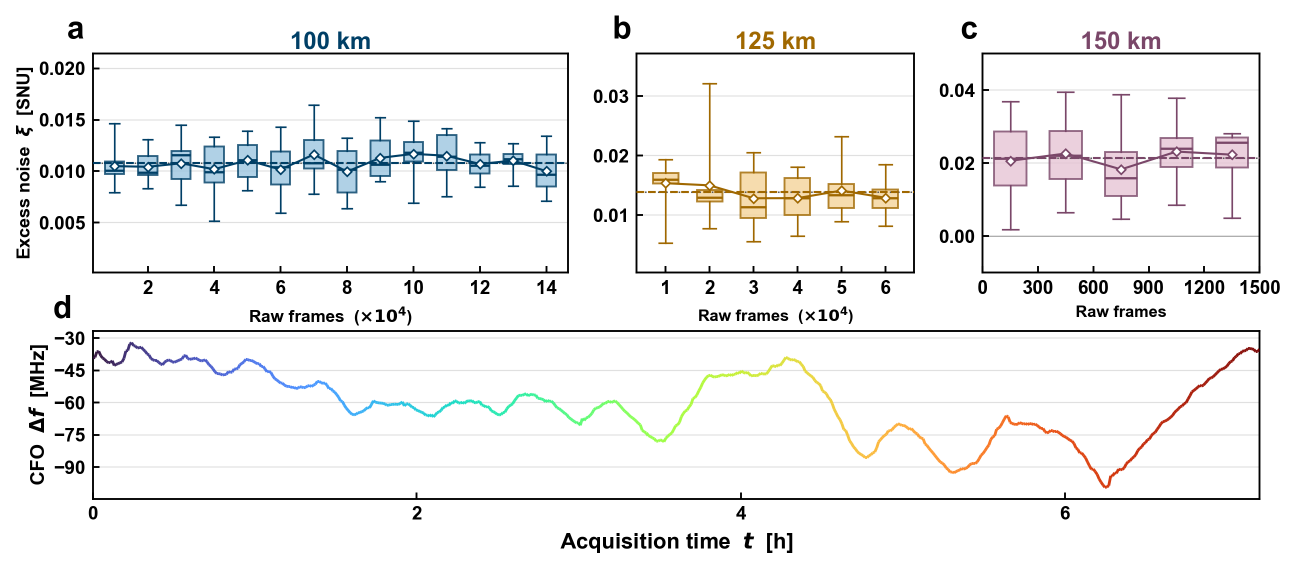}
\caption{\textbf{Excess-noise measurements and carrier-frequency tracking.} (a)--(c) Input-referred excess noise $\xi$ over acquisition blocks at 100, 125, and 150~km. (d) Pilot-estimated CFO, defined as the frequency difference between Alice's and Bob's free-running local-oscillator laser during the 100~km acquisition.}
\label{fig:longterm-stability}
\end{figure}

Achieving stable key generation over 100~km requires the optical front end, acquisition chain, and post-processing to function as a coupled system. To implement this, integrated lasers supply stable, narrow-linewidth carriers for Alice and Bob, while the silicon transmitter maintains calibrated quadrature modulation at the selected $V_A$. At the receiving end, the coherent receiver provides a flat response across the quantum band. For the digital processing, public pilots enable estimation of the carrier-frequency offset (CFO) and recovery of the residual carrier phase, while segmented acquisition forms the blocks used for security analysis. Subsequently, real-time shot-noise calibration maintains the SNU reference, and multidimensional reconciliation and privacy amplification extract secret keys from the low-SNR measurements. Detailed optical, electrical, and acquisition configurations are provided in Section~\ref{sec:experimental-setup} and Supplementary Note~6, whereas receiver DSP and key extraction are detailed in Sections~\ref{sec:phase-recovery} and~\ref{sec:postproc}, respectively. Figure~\ref{fig:system-dsp} maps this end-to-end workflow, from on-chip signal generation and coherent acquisition to receiver DSP, security analysis, and final key extraction. Figure~\ref{fig:longterm-stability} relates the stability of the integrated lasers to the phase-recovery performance and the blockwise excess-noise estimates. The ultra-narrow laser linewidths constrained phase-diffusion variance to $2.51 \times 10^{-6}~\text{rad}^2/\text{symbol}$ at 100~\text{MBaud}. Over 7.2~\text{hours}, the CFO stayed within 70~\text{MHz} with zero phase slips and a maximum drift of 4.59~\text{MHz} between adjacent 500-frame windows, enabling long-term acquisition over 100~\text{km}.

\begin{figure}[!htbp]
\centering
\includegraphics[width=0.7\linewidth]{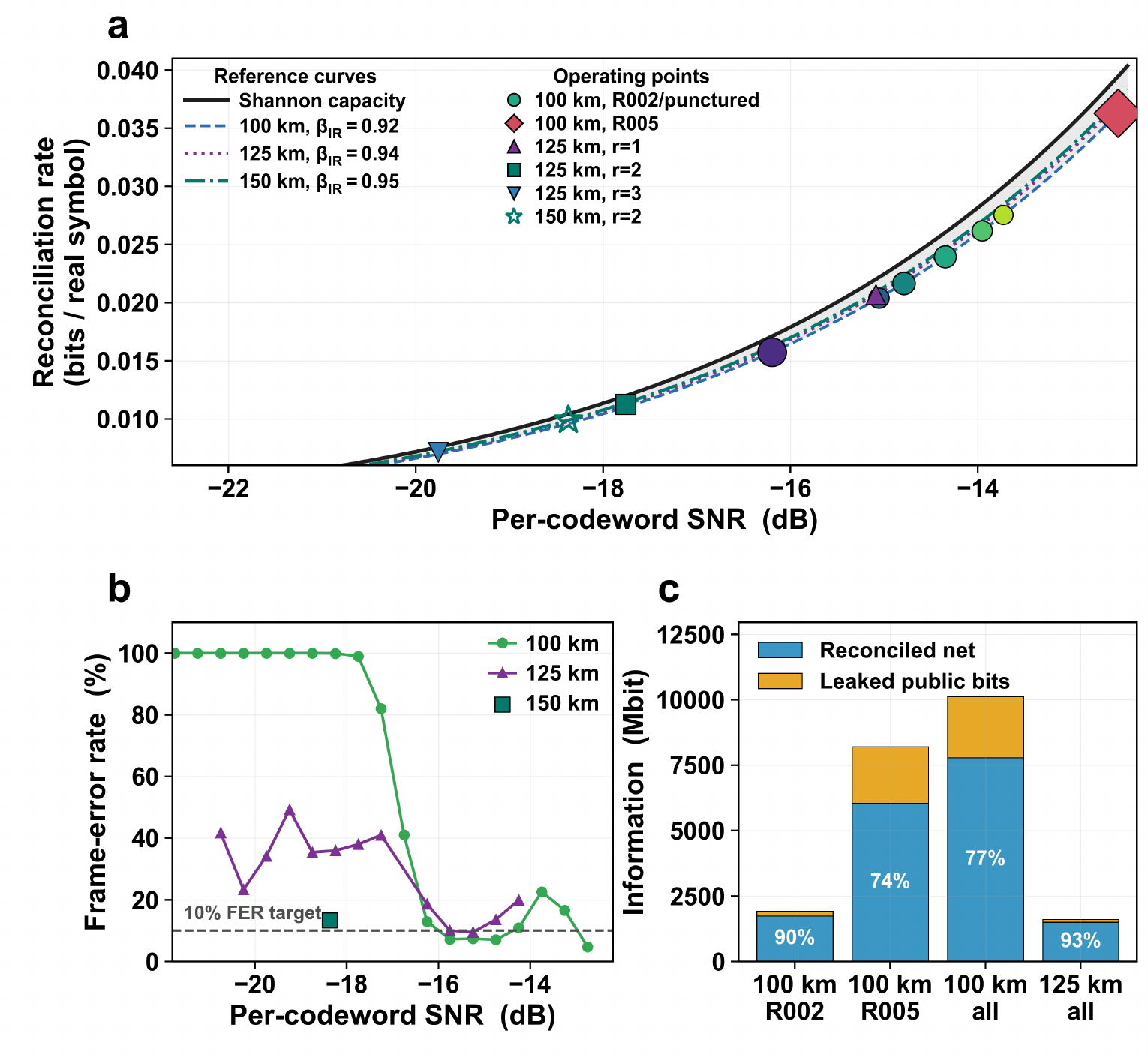}
\caption{\textbf{SNR-adaptive reconciliation.} (a) Reconciliation rate as a function of the publicly disclosed per-codeword SNR. (b) FER after public-coordinate disclosure at $\beta=0.92$, $0.94$ and $0.95$. (c) Verified information bits and public disclosure used in the 100 and 125~km reconciliation analysis.}
\label{fig:code-coverage}
\end{figure}

At 100~km, a block of $N=1.4\times10^{11}$ symbols acquired at $V_A=6.77$~SNU was used for parameter estimation in the composable finite-size analysis. The record was partitioned into 140 blocks of $10^9$ symbols for blockwise excess-noise estimation, giving a mean input-referred excess noise of $1.078\times10^{-2}$~SNU and a maximum of $1.64\times10^{-2}$~SNU.  We rounded this measured maximum upwards and used $\xi_{\mathrm{PE}}^U=1.65\times10^{-2}$~SNU in the Holevo bound. The block was then processed by multidimensional reverse reconciliation~\cite{LeverrierMultidim2008,Jouguet2013LongDistance} using rate-matched multi-edge-type low-density parity-check (MET-LDPC) codes. Codewords were routed according to the publicly disclosed per-codeword SNR, with the mother code, puncturing mask, and coordinate disclosures fixed before decoding. The measured reconciliation efficiency was $\beta=0.92$, with a frame error rate (FER) of 6.98\% (Fig.~\ref{fig:code-coverage}). The corresponding error-correction success probability, $p_{\mathrm{ec}}=0.93$, was used in the key-rate calculation. Using the measured excess-noise bound, the full-record transmittance confidence interval, and the nominal trusted receiver parameters $\eta_D=0.408$ and $v_{\mathrm{el}}=0.28$, with their worst-case drift adjustments applied as specified in Supplementary Note~1, we calculated the composable finite-size key with the improved Gaussian-modulation bound~\cite{PP2024ImprovedComposable}. Defining $K_{\mathrm{comp}}=\ell/N$ for the composable secret-key rate per channel use, the 100-km key-rate equation is expressed as:

\begin{equation}
K_{\mathrm{comp}}
=
\frac{1}{N}
\left\{
p_{\mathrm{ec}}
\left[
n\bigl(\beta I_{AB}-\chi_{BE}\bigr)
-\sqrt{n}\,\Delta_{\mathrm{AEP}}
+\theta
-1
\right]
-\mathrm{leak}_{\mathrm{PE}}
-n\lambda_{\mathrm{SC}}
\right\}.
\label{eq:keylen}
\end{equation}

Here we apply the improved composable bound of Ref.~\cite{PP2024ImprovedComposable} under the no-sacrifice parameter-estimation convention ~\cite{Lupo2018PE}, such that $n=N$. Decoding failures were included through $p_{\mathrm{ec}}$, while the information revealed during reconciliation was incorporated into $\beta$ (Methods and Supplementary Note~2). The finite-size terms $\Delta_{\mathrm{AEP}}$, $\theta$, $p_{\mathrm{ec}}$, $\mathrm{leak}_{\mathrm{PE}}$, and $\lambda_{\mathrm{SC}}$ are evaluated within the composable-security analysis detailed in Supplementary Note~1 prior to privacy amplification.  For the 100-km block, the guaranteed composable finite-size secret-key rate was $2.93\times10^{-4}$~bit/symbol at $\epsilon\le6\times2^{-32}$.

\begin{figure}[!htbp]
\centering
\includegraphics[width=0.7\linewidth]{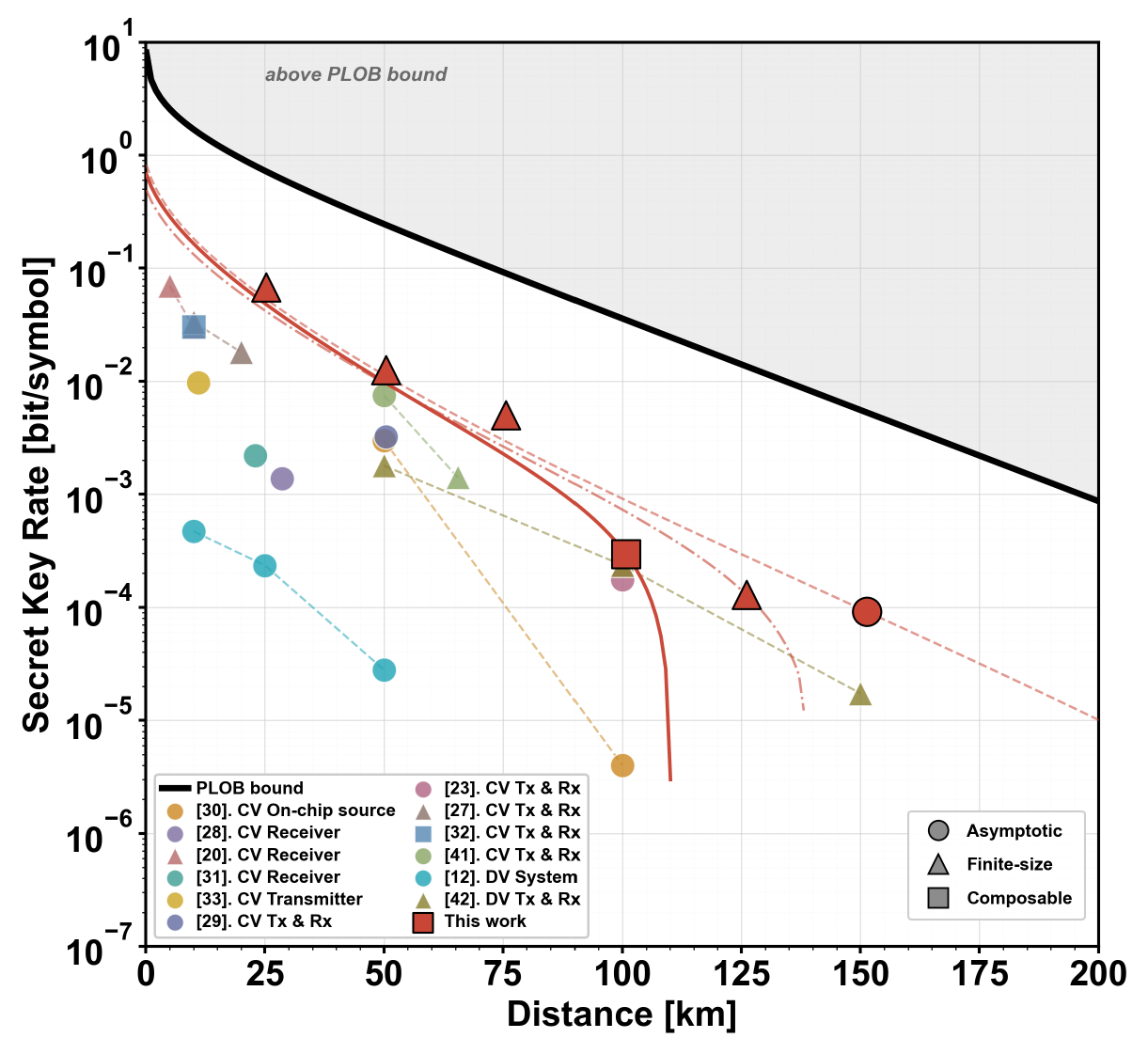}
\caption{\textbf{Secret-key rates of integrated QKD systems.} Secret-key rate per symbol for this work and previous chip-based QKD demonstrations~\cite{chip5,chip2,chip8,Paraiso2021,Wei2023,chip3,chip6,chip7,chip9,chip10,chip13,Xu2026Bidir}.}
\label{fig:skr-distance}
\end{figure}

For the 125-km block of $N=5.9\times10^{10}$ symbols acquired at $V_A=6.5$~SNU, the measured input-referred excess noise was $1.39\times10^{-2}$~SNU.  Each MET-LDPC codeword followed an SNR-adaptive soft-combining route selected based on the publicly disclosed channel state.  After normalisation by all physical channel uses consumed by these routes, the measured reconciliation efficiency was $\beta=0.94$, with an FER of 17.64\% (Fig.~\ref{fig:code-coverage}).  The finite-size secret-key rate was $1.29\times10^{-4}$~bit/symbol. Relative to verified single-observation batches, SNR-adaptive observation combining increased the secret-key rate by 3.13~kbps (32.0\%) (Supplementary Note~2). For the 150-km symbol block of $N=1.5\times10^8$ symbols acquired at $V_A=11$~SNU, the measured input-referred excess noise was $\xi=2.14\times10^{-2}$~SNU. Joint processing of soft information from independent observations before rate-matched MET-LDPC decoding achieved $\beta=0.95$, with an FER of 13.33\%.  After accounting for the measured FER, the resulting asymptotic secret-key rate was $9.17\times10^{-5}$~bit/symbol, corresponding to 9.17~kbps at 100~MBaud. The soft-information construction, code families, SNR windows, and per-distance routing statistics are given in Section~\ref{sec:postproc}, Supplementary Note~2, and Supplementary Tables~S1 and S2. Figure~\ref{fig:skr-distance} compares these rates with previous chip-based QKD demonstrations.

\section{Discussion}
We have demonstrated a CV-QKD link in which the chip light source, state preparation, and coherent reception constitute a single integrated hardware chain and a composable secret key has been certified directly from the data acquired by this platform. The 100-km system-level validation is therefore not a projection from optical characterization or an assumed reconciliation efficiency, but a validated system-level operating point at which the complete key-distillation chain is executed. The measured channel statistics, calibrated receiver parameters, finite block size, and achieved reconciliation performance are all rigorously incorporated into the composable finite-size analysis.

At 100~km, the system achieves a composable secret-key rate of \(2.93\times10^{-4}\)~bit/symbol, representing the highest reported per-symbol rate among state-of-the-art chip-based CV- and DV-QKD demonstrations. While several representative studies have established important benchmarks for integrated QKD, including resource-efficient DV-QKD with silicon-photonic encoder and decoder chips~\cite{Wei2023}, long-distance asymptotic TLO CV-QKD using an off-chip laser~\cite{chip2}, and high-speed composable key generation with integrated CV-QKD over approximately 10~km~\cite{chip10}, the present system builds on these advances to extend the demonstrated reach of composably secure integrated CV-QKD by a full order of magnitude. Beyond this operating point, the same platform successfully generates secret keys at 125~km under finite-size analysis and at 150~km under asymptotic analysis---setting the longest reported reaches for integrated CV-QKD in their respective security regimes.

These results further advance the realisation of a fully packaged integrated terminal. The hybrid III--V/Si$_3$N$_4$ laser source, silicon transmitter, and silicon receiver are compatible with CMOS integration, and the analogue front end can be co-designed with carrier recovery, reconciliation, and privacy-amplification electronics, in line with recent high-symbol-rate chip CV-QKD receivers~\cite{chip5,chip3}. In the present setup, the EDFA gain is mainly used to compensate the large fibre-to-chip coupling losses and chip packaging losses. Future integration on a single chip would eliminate much of this loss. If additional gain is still required, it could be supplied by an integrated InP BOA rather than a fibre amplifier, as shown in our previous work~\cite{Li:24}. Crucially, the intrinsic spectral filtering provided by coherent detection offers strong immunity against classical in-band noise, enabling seamless coexistence with telecom WDM infrastructure in standard single-mode fibres~\cite{120km}. The architecture thus enables a pluggable, high-density quantum-access terminal for practical deployment, while completely eliminating the cryogenic infrastructure required by high-efficiency SNSPD links~\cite{Li2025SNSPD99}.

The present optical reach is set by a system-level trade-off among overall detection efficiency, modulation variance, and ultralow-SNR reconciliation. Fibre-to-chip and inter-chip coupling losses, receiver detection loss, and trusted electronic noise reduce the received SNR and narrow the finite-size parameter-estimation margin. At 150~km, we therefore preselected a higher modulation variance to sustain reconciliation. This improves decoder operation but also amplifies modulation-dependent excess noise, particularly RPN, leaving less finite-size margin at the available block length. Even with the higher modulation variance, the available MET-LDPC codes require independent-observation combining, which sums LLRs but consumes multiple channel uses per decoded payload. Lower-loss optical interfaces, higher-responsivity photodiodes, and lower-noise receiver electronics could increase the end-to-end detection efficiency and recover part of the SNR margin~\cite{chip5,chip6,chip9}. Lower-rate, rate-compatible MET-LDPC ensembles can extend the positive-key SNR range~\cite{Milicevic2018QCMET,Jeong2022RateCompatible}. These changes target the limits observed here and could extend composable integrated CV-QKD beyond 100~km in practical quantum access networks.

\section{Methods}\label{sec:methods}
\subsection{Experimental setup }\label{sec:experimental-setup}

The experimental link comprised two integrated laser modules, a silicon transmitter module at Alice, and a silicon coherent receiver module at Bob, as shown in Fig.~\ref{fig:system-dsp}. Alice's waveform was prepared from two independent Gaussian quadrature sequences at the 100~MBaud protocol rate and five samples per symbol. After root-raised-cosine (RRC) pulse shaping, the sequences were shifted to a single-sideband subcarrier, combined with a continuous-wave pilot tone, and pre-emphasized to compensate the measured electro-optic frequency response across the quantum band. The measured pre-emphasis response and spectra before and after correction are shown in Fig.~\ref{fig:system-dsp}(b).

In the optical path, an erbium-doped fibre amplifier (EDFA) amplified the integrated laser output from about $-1$ to $13$~dBm before modulation. An integrated variable optical attenuator (VOA) then set the modulation variance, and fibre polarization controllers aligned the optical states at the module interfaces. The channel was assembled from Corning SMF-28e ULL fibre and YOFC ULL-LAF to form the 25.25 to 151.32~km distance ladder of Table~\ref{tab:sota}, with an effective attenuation of $0.161$~dB/km measured by an optical time-domain reflectometer (OTDR), including connector loss. At Bob, a second integrated laser generated the local oscillator. Under the same receiver configuration, the optical switch alternated between the quantum and vacuum windows to acquire shot-noise traces as described in Section~\ref{sec:snu}. The oscilloscope and sustained-acquisition method are detailed in Supplementary Note~6.

\subsection{Phase recovery}\label{sec:phase-recovery}
At low SNRs over long links, residual carrier-phase error directly adds to excess noise. Conventional forward-only phase recovery~\cite{Shao2021PhaseNoise} conditions estimates solely on past samples, leaving subsequent pilot observations unexploited. To leverage the full data record, we augment the forward UKF with a Rauch--Tung--Striebel (RTS) backward pass~\cite{Rauch1965RTS}. The resulting unscented RTS (URTS) smoother combines stored forward-pass moments with later observations via a single backward recursion, preserving the interpretable phase-frequency state-space model of the UKF. We represented the residual dynamics with a phase-and-frequency state and a unit-phasor observation model,

\begin{equation}
\theta_{k+1}=\theta_k+\omega_k\,\Delta t+w_{\theta,k},\qquad \omega_{k+1}=\omega_k+w_{\omega,k},\qquad \mathbf{z}_k=\big(\cos\theta_k,\ \sin\theta_k\big)^{\!\top}+\mathbf{n}_k.
\label{eq:methods-phase-state}
\end{equation}

Here, $\theta_k$ is the residual carrier phase, $\omega_k$ is the residual angular-frequency offset, and $\mathbf{z}_k$ is the two-dimensional real representation of the unit phasor. The process-noise vector $\mathbf{w}_k=(w_{\theta,k},w_{\omega,k})^{\top}$ describes unmodelled phase and frequency evolution between pilot observations, including slow relative drift. The vector $\mathbf{n}_k$ denotes the corresponding observation noise. The process and observation noises were assumed to be zero-mean, with covariances $Q$ and $R$, respectively. The UKF sigma points were generated using a scaled unscented transform and the covariance matrices were set to $Q=\operatorname{diag}(10^{-6}~\mathrm{rad}^2,\, 10^{8}~\mathrm{rad}^2\,\mathrm{s}^{-2})$ and $R=0.3\mathbf{I}_2$. On the decimated pilot grid, the forward UKF generated the filtered state estimates, predicted moments, and covariances required by the RTS backward recursion. The backward pass yielded the smoothed phase trajectory $\hat{\theta}^{\mathrm{RTS}}_k$ and covariance $\bm{P}^{\mathrm{RTS}}_{k|N}$, conditioned on all $N$ pilot observations in the record. The smoothed phase trajectory was then interpolated to the full sampling grid and applied to the corresponding received-waveform samples.

To isolate the contribution of smoothing, the UKF and URTS used the same phase-frequency model and fixed covariance matrices. Source acquisitions were randomly sampled, and the fifth cut frame from each source was fixed before either estimator was run. After common coarse-frequency recovery and pilot extraction, the first 20 decimated pilot observations initialized the shared state, and every fifth subsequent observation was masked. At masked indices, the UKF propagated the state without a measurement update; the URTS backward pass used only the stored forward states and covariances from the same masked run together with the fixed process covariance. Circular phase error was evaluated against the masked observations after both trajectories were complete.

\subsection{Real-time shot-noise calibration and acquisition}\label{sec:snu}
An OS, driven by a square wave from an arbitrary waveform generator, alternated between quantum and vacuum acquisitions under identical receiver settings. Each vacuum trace defined the SNU for the adjacent quantum window after the same receiver DSP. Each key-record acquisition cycle comprised 2~s of quantum-signal acquisition followed by 1~s of vacuum acquisition. The vacuum traces were processed using the same whitening convention, DSP operations, and normalisation as the quantum traces, thereby referring the SNU calibration to the same detected receiver mode~\cite{Sun2025ContinuousMode}. Shot-noise stability was assessed by the Allan-deviation analysis in Fig.~S1. The panel of Fig.~\ref{fig:system-dsp}(b) shows the electronic-noise, shot-noise, and 100~km quantum spectra after noise-floor matching.

\subsection{Reverse reconciliation}\label{sec:postproc}
The normalised post-DSP quadrature streams were grouped into eight-dimensional vectors $\bm{x}_j$ and $\bm{y}_j$ for multidimensional reverse reconciliation~\cite{LeverrierMultidim2008,Jouguet2013LongDistance}. Bob sampled a uniformly random binary string $\bm{c}\in\{0,1\}^{N_{\mathrm{c}}}$ and mapped its bits as $u_t=1-2c_t$, forming $\bm{u}_j\in\{\pm1\}^{8}$. For each vector, Bob constructed a public orthogonal map from $\bm{y}_j/\lVert\bm{y}_j\rVert$ to $\bm{u}_j/\sqrt{8}$. Bob transmitted the map together with the MET-LDPC syndrome $S=H\bm{c}$. Alice applied the same map to $\bm{x}_j/\lVert\bm{x}_j\rVert$ and obtained the rotated unit vector $\bm{r}_j$. The corresponding binary-input LLRs were
\begin{equation}
L_{j,t}
= \frac{2\,r_{j,t}\,|x_j|\,|y_j|}{\sqrt{8}\,\sigma^2},
\label{eq:mdr-llr}
\end{equation}
where $t=1,\ldots,8$ and $r_{j,t}$ is the $t$th coordinate of $\bm{r}_j$. Here $|x_j|=\lVert\bm{x}_j\rVert$, $|y_j|=\lVert\bm{y}_j\rVert$, and $\sigma^2$ is the residual quadrature-noise variance inferred from the public SNR. Equation~\eqref{eq:mdr-llr} describes the binary-input AWGN channel induced by the orthogonal mapping~\cite{LeverrierMultidim2008,Milicevic2018QCMET}. The LLR sign gives the provisional bit value, while its magnitude sets the reliability used by the belief-propagation decoder.

The per-codeword SNR varied across the acquisition campaign, so the reconciliation route was selected dynamically per codeword. The decoder used MET-LDPC code families with base rates of 0.01995 and 0.05~\cite{Mani2021METLDPC}. Puncturing and coordinate disclosure set the effective code rate~\cite{Wang2017RateAdaptive,MartinezMateo2012Blind,Gumus2021MDA}. Supplementary Note~2 gives the SNR windows, puncturing masks, and realised rates.

The target $\beta$ and selected route fixed the disclosure count $s$ before belief-propagation decoding. Coordinates were taken from the public puncture mask when possible, with any additional positions selected by reliability and reported explicitly~\cite{Gumus2021MDA}. The corresponding LLRs were replaced by fixed-magnitude values whose signs matched the disclosed bits. Offline success required zero syndrome mismatch and exact agreement with the reference string. A $t_V$-bit universal-hash tag provides correctness verification in deployment~\cite{ref42}. To maintain a strict upper bound on information leakage, the maximum planned disclosure was uniformly charged to every codeword, regardless of whether decoding converged without additional disclosure. Supplementary Note~2 gives the complete disclosure rule and public-message accounting.

At the lowest SNRs, LLRs from independent physical observations were combined before a single MET-LDPC decoding attempt. To guarantee valid finite-size accounting, all constituent observations and their corresponding post-DSP frames were strictly retained in the total physical channel-use count. Each combined group produced at most one decoded binary string. Supplementary Note~2 states the conditional-independence assumptions and the corresponding Holevo-information accounting.

Total reconciliation leakage was calculated from all public messages, including syndromes, disclosed coordinates and values, failure messages, and verification tags. The residual correctness failure probability entered separately through $\theta$. The verified blocks were concatenated and compressed with a universal Toeplitz hash~\cite{ref42} to the length specified by Eq.~\eqref{eq:keylen}. All public communication was information-theoretically authenticated using a short pre-shared key replenished from the generated key material.

\paragraph{Funding.} This work is supported by National Natural Science Foundation of China (No. 62571316, 61971276), Quantum Science and Technology-National Science and Technology Major Project (No. 2021ZD0300703), Shanghai Municipal Science and Technology Major Project (No. 2019SHZDZX01), Natural Science Foundation of Shanghai (No. 25ZR1402251), and Cultivation Project of Shanghai Research Center for Quantum Sciences (No. LZPY2024).

\paragraph{Data availability.} 	All of the data that support the findings of this study are reported in the main text and the Supplementary Information. Source data are available from the corresponding authors on reasonable request.

\paragraph{Acknowledgments.} The authors thank the team at Sifotonics for the collaboration on the PIC design and for fabricating the chips. The computations in this paper were partly run on the $\pi$ 2.0 cluster supported by the Center for High Performance Computing at Shanghai Jiao Tong University.

\paragraph{Competing interests.} The authors declare no competing interests.

\paragraph{Author contributions statement.} G.Z. conceived and supervised the research. Y.X. and X.L. performed the theoretical derivation and experimental analysis. Y.X. and X.L. analysed the data and wrote the manuscript. Y.X., T.W., X.L., X.J., Y.G., L.Z. and P.H. provided technical support for the experimental system and data collection. All authors discussed the results and reviewed the manuscript.

\bibliographystyle{naturemag-doi}
\bibliography{ref}

\end{document}


\begin{center}
{\LARGE\bfseries Supplementary Information\par}
\vspace{1.25em}
{\Large\bfseries Fully integrated continuous-variable quantum key distribution with composable security over 100 km\par}
\vspace{1em}
{\normalsize Yankai Xu, Xinhang Li, Tao Wang, Jisheng Dai, Xueqin Jiang, Yuyao Guo, Peng Huang, Linjie Zhou, and Guihua Zeng\par}
\vspace{0.8em}
{\small
State Key Laboratory of Photonics and Communications, Shanghai Jiao Tong University, Shanghai 200240, China\\
Shanghai Research Center for Quantum Sciences, Shanghai 201315, China\\
Hefei National Laboratory, Hefei 230088, China\\
SJTU-Pinghu Institute of Intelligent Optoelectronics, Pinghu 314200, China\\
College of Information Science and Technology, Donghua University, Shanghai 201315, China\\[3pt]
Correspondence: tonystar@sjtu.edu.cn; ghzeng@sjtu.edu.cn
\par}
\end{center}

\vspace{0.8em}
\hrule
\vspace{1.2em}
\section{Asymptotic, finite-size, and composable GMCS}\label{sec:app-gmcs-accounting}

Alice's modulation variance is $V_A$, with entanglement-based variance $V=V_A+1$ in shot-noise units (SNU).  The untrusted parameters are the transmittance $T$ and input-referred excess noise $\xi$; the trusted receiver parameters are the detection efficiency $\eta_D$ and per-quadrature electronic noise $v_{\mathrm{el}}$.  We evaluate $I_{AB}$ on the estimated channel and maximize $\chi_{BE}$ over the finite-size confidence region. The noise terms are $\chi_{\mathrm{line}}=(1-T)/T+\xi=\xi-1+1/T$, $\chi_{\mathrm{het}}=(2-\eta_D+2v_{\mathrm{el}})/\eta_D$, and $\chi_{\mathrm{tot}}=\chi_{\mathrm{line}}+\chi_{\mathrm{het}}/T$.  In the composable finite-size analysis~\cite{PP2024ImprovedComposable}, reverse-reconciled heterodyne detection gives the Alice--Bob signal-to-noise ratio (SNR) $\mathrm{SNR}_{AB}=V_A/(1+\chi_{\mathrm{tot}})$ and mutual information $I_{AB}=\log_2(1+\mathrm{SNR}_{AB})$.

Before trusted detection, the covariance matrix is $\gamma_{AB}=\left(\begin{smallmatrix}V\mathbb{I}_2&\sqrt{T(V^2-1)}\sigma_z\\\sqrt{T(V^2-1)}\sigma_z&[T(V+\xi)+1-T]\mathbb{I}_2\end{smallmatrix}\right)$, where $\sigma_z=\mathrm{diag}(1,-1)$.  The four symplectic invariants are~\cite{theo2} $\mathcal A=V^2(1-2T)+2T+T^2(V+\chi_{\mathrm{line}})^2$ and $\mathcal B=T^2(1+V\chi_{\mathrm{line}})^2$, together with $\mathcal C=[\mathcal A\chi_{\mathrm{het}}^2+\mathcal B+1+2\chi_{\mathrm{het}}(V\sqrt{\mathcal B}+T(V+\chi_{\mathrm{line}}))+2T(V^2-1)]/[T^2(V+\chi_{\mathrm{tot}})^2]$ and $\mathcal D=[(V+\sqrt{\mathcal B}\chi_{\mathrm{het}})/(T(V+\chi_{\mathrm{tot}}))]^2$. The unconditional and nontrivial conditional eigenvalues are $\lambda_{1,2}=\sqrt{[\mathcal A\pm\sqrt{\mathcal A^2-4\mathcal B}]/2}$ and $\lambda_{3,4}=\sqrt{[\mathcal C\pm\sqrt{\mathcal C^2-4\mathcal D}]/2}$.  The Holevo quantity is $\chi_{BE}=G[(\lambda_1-1)/2]+G[(\lambda_2-1)/2]-G[(\lambda_3-1)/2]-G[(\lambda_4-1)/2]$, where $G(x)=(x+1)\log_2(x+1)-x\log_2 x$.  The unconditional eigenvalues depend on $(T,\xi)$, whereas $(\eta_D,v_{\mathrm{el}})$ enter the conditional eigenvalues through $\chi_{\mathrm{het}}$.  The remaining conditional eigenvalue is unity and contributes zero.  Gaussian optimality makes this $\chi_{BE}$ an upper bound for collective attacks with the estimated covariance~\cite{GarciaPatron2006Optimality,Navascues2006Optimality}.  The asymptotic reverse-reconciled rate is $K_{\infty}=\beta I_{AB}-\chi_{BE}$, where $\beta$ denotes the reconciliation efficiency.

For finite-size accounting, $N$ denotes the number of prepared states and $n$ the reconciliation length.  Under no-sacrifice parameter estimation, the same record supports both tasks~\cite{Lupo2018PE}.  At 100~km, $n=N=1.40\times10^{11}$ and $m_{\mathrm{eff}}=N$, where $m_{\mathrm{eff}}$ counts reused estimator samples. For heterodyne detection the transmittance confidence interval follows from the estimator variance.
\begin{equation}
\sigma_T
=\frac{2\hat T}{\sqrt{2m_{\mathrm{eff}}}}
\sqrt{
c_{\mathrm{PE}}
+\frac{\hat\xi+\bigl(2+2v_{\mathrm{el}}\bigr)/(\eta_D\hat T)}{V_A}
},
\qquad
T_{L,U}=\hat T\mp w\sigma_T,
\qquad
w=\sqrt{2}\operatorname{erf}^{-1}(1-\epsilon_T).
\label{eq:app-transmittance-ci}
\end{equation}
Calculations use unrounded inputs; decimal values are reported to three significant figures.  We use $c_{\mathrm{PE}}=0$, $V_A=6.77$, the drift-adjusted values $\eta_D=0.404$ and $v_{\mathrm{el}}=0.285$, and $\hat\xi=1.08\times10^{-2}$.  For $m_{\mathrm{eff}}=N$, $\hat T=2.38\times10^{-2}$, and $\epsilon_T=2^{-32}$, Eq.~\eqref{eq:app-transmittance-ci} gives $w\sigma_T/\hat T=1.50\times10^{-4}$. The corresponding analytic excess-noise is $1.543\times10^{-2}$~SNU~\cite{PP2024ImprovedComposable}. Comparing it with the largest blockwise estimate, $1.6423\times10^{-2}$~SNU, and rounding the larger value upwards gives the final independent bound $\xi_{\mathrm{PE}}^U=1.65\times10^{-2}$~SNU with $\epsilon_\xi=2^{-32}$. Because the experimentally determined excess-noise upper bound $\xi_{\mathrm{PE}}^U$ is imposed independently, we fix $\xi=\xi_{\mathrm{PE}}^U$ and maximize the Holevo term over the transmittance confidence interval.
\begin{equation}
\chi_{BE}^{U}
=\max_{T\in[T_L,T_U],\,0\le\xi\le\xi_{\mathrm{PE}}^U}
\chi_{BE}(T,\xi)
=\chi_{BE}(T_U,\xi_{\mathrm{PE}}^U).
\label{eq:app-holevo-confidence-max}
\end{equation}
For the 100~km block, direct evaluation shows that the maximum occurs at $T_U$, giving $\chi_{BE}^{U}=\chi_{BE}(T_U,\xi_{\mathrm{PE}}^U)$. The collective-attack calculation uses the following union-bound failure budget.
\begin{equation}
\epsilon_{\mathrm{tot}}
\le
\epsilon_{\mathrm{corr}}
+\epsilon_s
+\epsilon_h
+\epsilon_T
+\epsilon_\xi
+\epsilon_{\mathrm{auth}}.
\label{eq:epsilon-budget}
\end{equation}
The terms cover correctness, smoothing, hashing, transmittance estimation, excess-noise estimation, and authentication.  Setting each to $2^{-32}$ gives $\epsilon_{\mathrm{tot}}\le6\times2^{-32}\approx1.40\times10^{-9}$. Equation~(2) uses this composable finite-size lower bound, including its $-1$-bit achievable-length correction~\cite{PP2024ImprovedComposable}.  We use $\beta=0.920$ and $\theta=\log_2(2\epsilon_h^2\epsilon_{\mathrm{corr}})=-95$~bits.  Only verified codewords contribute to the reconciled payload; decoding failures enter once through the conservative prefactor $p_{\mathrm{ec}}=0.93$.  The reported value is the gross rate before session-level authentication-key replenishment. The rate calculation charges transmitter back-reflection side-channel (SC) leakage using $A_{\mathrm{iso}}\ge80.0$~dB.  This gives $\lambda_{\mathrm{SC}}=G(\tfrac12\,10^{-A_{\mathrm{iso}}/10}V_A)\le9.20\times10^{-7}$ bit/symbol and $\mathrm{leak}_{\mathrm{SC}}=n\lambda_{\mathrm{SC}}\le1.30\times10^{5}$ bits. The asymptotic equipartition property (AEP) finite-size entropy term is~\cite{ref17,ref18,PP2024ImprovedComposable}
\begin{equation}
\Delta_{\mathrm{AEP}}
=4\log_2\!\bigl(\sqrt{|L|}+2\bigr)
\sqrt{-\log_2\!\left(1-\sqrt{1-\epsilon_s^2}\right)}.
\label{eq:app-delta-aep}
\end{equation}
With $|L|=2^{14}$, $\epsilon_s=2^{-32}$, and the parameters above, Eq.~(2) gives $K_{\mathrm{comp}}=2.93\times10^{-4}$~bit/symbol.

\section{Reconciliation}\label{sec:app-reconciliation}
The reconciliation efficiencies were 0.92, 0.94, and 0.95 at 100, 125, and 150~km, respectively, and set the disclosure count through Eq.~\eqref{eq:postproc-beta}.  Failed codewords enter the frame error rate (FER), while only codewords that pass decoding and the syndrome check contribute verified information bits.  For a puncturing mask with $p$ variable nodes, the channel log-likelihood ratios (LLRs) of those nodes are set to zero and the effective block length becomes $N_{\mathrm{c}}-p$.  With $k$ information bits, single-observation capacity per real dimension $C_{\mathrm{avg}}=\tfrac12\log_2(1+\mathrm{SNR})$, and observation-combining factor $r_{\mathrm{comb}}$, the route-ladder rate and public-coordinate disclosure rule for the reported $\beta$ are
\begin{equation}
R_{\mathrm{route}}
=\frac{k-s}{r_{\mathrm{comb}}(N_{\mathrm{c}}-p)-s},
\qquad
\beta_{\mathrm{route}}=\frac{R_{\mathrm{route}}}{C_{\mathrm{avg}}},
\qquad
s=\left\lfloor
\frac{k-\beta C_{\mathrm{avg}}\,r_{\mathrm{comb}}(N_{\mathrm{c}}-p)}
{1-\beta C_{\mathrm{avg}}}
\right\rfloor .
\label{eq:postproc-beta}
\end{equation}
Here $R_{\mathrm{route}}$ is used for route selection, and $s$ is clipped to $0\le s\le k$.  The disclosed values act as shortening information~\cite{Wang2017RateAdaptive}.  All observations, including punctured and disclosed coordinates, remain included in $n$ and in the Holevo term.

Let $\mathcal V$ denote the codewords that passed decoding and the syndrome check.  For codeword $i$, let $e_i=(s_i-p_i)_+$ and $\ell_{P,i}=\lceil\log_2\binom{N_{\mathrm c}-p_i}{e_i}\rceil$, with $\ell_{P,i}=0$ when $e_i=0$. After syndrome disclosure, each retained codeword contains $k_i$ information bits.  Disclosed values subtract $s_i$ bits, and positions outside the public puncture mask subtract $\ell_{P,i}$ bits~\cite{Gumus2021MDA}; positions fixed by the mask require no position message.  A single $t_V=32$-bit tag is charged once and gives $\epsilon_{\mathrm{corr}}=2^{-32}$ for the concatenated data~\cite{ref42}.

At 100~km, we used rate-0.01995 and rate-0.05 multi-edge-type low-density parity-check (MET-LDPC) codes with $N_{\mathrm{c}}=10^6$ and $k=19{,}950$ or $50{,}000$, respectively.  Both use the multi-edge-type degree distributions of Ref.~\cite{Mani2021METLDPC} and quasi-cyclic liftings with circulant sizes 50 and 20.  The route table was fixed before decoding; the public SNR assigned each codeword to a mother code and puncture mask, and Eq.~\eqref{eq:postproc-beta} set any additional disclosure.  The fp32 decoder used flooding-schedule sum-product updates, with syndrome checks every five iterations and at most 500 iterations~\cite{Milicevic2018QCMET}.

\begin{table}[htbp]
\centering
\scriptsize
\setlength{\tabcolsep}{4pt}
\begin{threeparttable}
\caption{\textbf{The public SNR determines the reconciliation route at 100~km.}  The realised route rate includes the disclosure set by Eq.~\eqref{eq:postproc-beta}.}
\label{tab:code-route-100km}
\begin{tabular}{llcc}
\toprule
Route & Public SNR window (dB) & Realised $R_{\mathrm{route}}$ \\
\midrule
R002    & $<-15.15$           & up to $0.020$ \\
R002p021            & $[-15.15,-14.96)$   & $0.020$ to $0.021$ \\
R002p0230          & $[-14.96,-14.56)$   & $0.021$ to $0.023$ \\
R002p0253          & $[-14.56,-14.11)$   & $0.023$ to $0.025$ \\
R002p0270          & $[-14.11,-13.82)$   & $0.025$ to $0.027$ \\
R002p0281          & $[-13.82,-13.63)$   & $0.027$ to $0.028$ \\
R005               & $\ge -13.63$        & $\ge 0.028$ \\
\bottomrule
\end{tabular}
\end{threeparttable}
\end{table}

For fixed public classes $Q^r=q^r$, the observations satisfy $\rho_{X^rY^rE^r\mid q^r}=\bigotimes_m\rho^{(m)}_{X_mY_mE_m\mid q_m}$ and retain the rotational symmetry required by the multidimensional map~\cite{LeverrierMultidim2008}.  Here $Q_m$ contains the SNR bin, route, puncture mask, and calibrated LLR parameters.  To ensure rigorous statistical independence, combination groups were fixed blindly prior to decoding, strictly prohibiting any reuse of observations. Conditional on $(\bm\alpha_m,q_m)$, the induced channels are modelled as memoryless binary-input symmetric channels.  The calibrated single-observation LLRs combine as~\cite{Chase1985CodeCombining}
\begin{equation}
L^{(\mathrm{comb})}_{j,t}
=\sum_{m=1}^{r_{\mathrm{comb}}} L_{m,j,t}.
\label{eq:llr-comb}
\end{equation}
All $r_{\mathrm{comb}}$ observations remain in the channel-use denominator, while each group produces at most one decoded information string. Ref.~\cite{LeverrierMultidim2008} proves that the public map for one isotropic observation is independent of the uniformly generated string.  Under this model, the single-map result extends to all maps in a combined group,
\begin{equation}
p(\bm\alpha^r\mid\bm u,q^r)
=\prod_{m=1}^{r}p(\bm\alpha_m\mid\bm u,q_m)
=\prod_{m=1}^{r}p(\bm\alpha_m\mid q_m).
\label{eq:joint-map-independent}
\end{equation}

\begin{table}[!htbp]
\centering
\scriptsize
\setlength{\tabcolsep}{2.6pt}
\begin{threeparttable}
\caption{\textbf{Measured reconciliation by distance and observation-combining factor.}  The factor $r_{\mathrm{comb}}$ is the number of separate observations combined per decoding attempt.}
\label{tab:reconciliation-transcript}
\begin{tabular}{lcrrrrc}
\toprule
Distance & $r_{\mathrm{comb}}$ & \makecell{Attempted\\blocks} & \makecell{Successful\\blocks} & FER & \makecell{Physical\\frames} & $\beta$ \\
\midrule
100~km & 1 & 280{,}000 & 260{,}459 & 6.98\% & 1{,}400{,}000 & 0.92 \\
\addlinespace
125~km & 1 & 81{,}800 & 70{,}639 & 13.64\% & 409{,}000 & \\
125~km & 2 & 11{,}837 & 7{,}316 & 38.19\% & 118{,}370 & \\
125~km & 3 & 4{,}244 & 2{,}663 & 37.25\% & 63{,}660 & \\
125~km & 1--3 & 97{,}881 & 80{,}618 & 17.64\% & 591{,}030 & 0.94 \\
\addlinespace
150~km & 2 & 150 & 130 & 13.33\% & 1{,}500 & 0.95 \\
\bottomrule
\end{tabular}
\end{threeparttable}
\end{table}

This relation allows the single-map Holevo bound to be applied to the combined group.  Applying the single-map independence result and Lemma~1 of Ref.~\cite{LeverrierMultidim2008} to the aggregate variables, followed by the standard additivity of Holevo information for a tensor-product cq state, gives
\begin{equation}
\chi(\bm u:E^r,\bm\alpha^r\mid q^r)
\leq \chi(\bm y^r:E^r\mid q^r)
=\sum_{m=1}^{r}\chi(\bm y_m:E_m\mid q_m).
\label{eq:multiobs-holevo-bound}
\end{equation}
The bound is evaluated for each public class, and a common value must upper-bound every class used by the route.  For i.i.d. observations, the sum reduces to $r\chi_{BE}$.  The combined observations came from distinct oscilloscope captures and had mean cross-frame coherence below 0.005 in a signal-free analysis band.

At 125~km, the public SNR selected $r_{\mathrm{comb}}\in\{1,2,3\}$ according to a fixed route table. The codeword FER, computed per decoding attempt, was 17.64\%. Because an order-$r$ route consumed $r$ physical observations but produced at most one verified string, the finite-size rate was scaled by the physical-use-weighted success factor $p_{\mathrm{ec}}^{(\mathrm{use})}=\sum_r rN_{\mathrm{succ},r}/\sum_r rN_{\mathrm{att},r}\approx0.789$. For the single-observation reference, frames assigned to the $r_{\mathrm{comb}}=2$ and 3 routes were treated as decoding failures, while all physical channel uses remained in the denominator. This yielded 9.77~kbps, compared with 12.9~kbps for the mixed-route scheme. At 150~km, a fixed $r_{\mathrm{comb}}=2$ route was used within the public-SNR window $[-18.40,-18.35)$~dB at $\beta=0.95$. The asymptotic rate used a success factor of 0.867, corresponding to an FER of 13.33\%.

\section{Excess-noise model}\label{sec:app-device-budget}

The excess noise decomposition is $\xi=\xi_{\mathrm{RIN}}+\xi_{\mathrm{mod}}+\xi_{\mathrm{RPN}}+\xi_{\mathrm{other}}$.  The term $\xi_{\mathrm{other}}$ contains additive leakage and unresolved cross-covariances after the receiver DSP.  It also includes analogue-to-digital converter (ADC) and receiver-digitisation residuals not already represented by the trusted $v_{\mathrm{el}}$, together with calibration drift and the small spontaneous Raman background.

\textbf{Relative intensity noise.} Let $g_a$ be a small fractional field-gain fluctuation in optical path $a$.  The linearized power fluctuation obeys $\delta P/P\simeq2g_a$, so quantum-band filtering gives~\cite{theo2,chip8}
\begin{equation}
\sigma^2_{g,a}
=
\frac{1}{4}
\int_{B_{\mathrm q}}
S_{\mathrm{RIN},a}(f)\,
\left|H_{\mathrm q}(f)\right|^2\,df .
\label{eq:rin-integral}
\end{equation}
Here $S_{\mathrm{RIN},a}(f)$ is the single-sided RIN spectrum in linear units, and $H_{\mathrm q}(f)$ is the normalised quantum-band response. For the signal arm, the gain fluctuation multiplies Alice's displacement and gives $\xi_{\mathrm{RIN,sig}}=V_A\sigma^2_{g,\mathrm{sig}}$.  The measured $-150$~dBc/Hz spectrum over $B_{\mathrm q}\simeq100$~MHz gives $\sigma^2_{g,\mathrm{sig}}=2.5\times10^{-8}$ and $\xi_{\mathrm{RIN,sig}}=1.69\times10^{-7}$~SNU at $V_A=6.77$.  The receiver-plane reference levels were $6.99\times10^{-4}$~SNU for ADC quantisation, $1.86\times10^{-4}$~SNU for finite common-mode rejection, and $5.47\times10^{-11}$~SNU for local oscillator (LO) RIN.

\textbf{Residual phase noise.} Let $\delta\phi_k=\phi_k-\hat\phi_k$ be the phase error after pilot recovery.  We model it as zero-mean Gaussian with variance $\sigma_\phi^2$~\cite{theo2,Shao2021PhaseNoise}. Let $T_0$ denote the optical transmittance before the residual rotation is absorbed into parameter estimation.  The mean quadrature gain contains $c=\mathbb E[\cos\delta\phi_k]=\exp(-\sigma_\phi^2/2)$.  Supplementary Note~\ref{sec:app-gmcs-accounting} therefore uses the fitted transmittance $T=T_0c^2$.  The residual signal variance per quadrature is $T_0V_A(1-c^2)$, which gives
\begin{equation}
\xi_{\mathrm{RPN}}
=
\frac{T_0V_A(1-c^2)}{T_0c^2}
=V_A(c^{-2}-1)
=V_A\left[\exp(\sigma_\phi^2)-1\right].
\label{eq:rpn-fitted-reference}
\end{equation}
For $\sigma_\phi^2\ll1$, Eq.~\eqref{eq:rpn-fitted-reference} reduces to $V_A\sigma_\phi^2$. The intrinsic white-frequency-noise contribution is treated explicitly as a separate linewidth-diffusion term. For a symbol period $\tau$, the two-laser contribution is
\begin{equation}
V_{\mathrm{drift}}
=
2\pi\left(\Delta\nu_{\mathrm{sig}}+\Delta\nu_{\mathrm{LO}}\right)\tau .
\label{eq:linewidth-drift}
\end{equation}
At 100~MBaud, $\tau=10$~ns and the measured intrinsic linewidths are $\Delta\nu_{\mathrm{sig}}=10$~Hz and $\Delta\nu_{\mathrm{LO}}=30$~Hz, giving $V_{\mathrm{drift}}=2.51\times10^{-6}$~rad$^2$, or $V_A V_{\mathrm{drift}}=1.70\times10^{-5}$~SNU at $V_A=6.77$.  Integrating the measured pilot line against its local noise floor over the $\pm1$~MHz tracking band gives $\mathrm{SNR}_p\simeq23$~dB at 100~km. For the accepted pilot samples $\mathcal A$, the experimental diagnostic uses the mean Rauch--Tung--Striebel (RTS) posterior phase variance~\cite{Rauch1965RTS}
\begin{equation}
V_{\mathrm{est}}
=
\frac{1}{N_p}\sum_{k\in\mathcal A}
\left[\bm P^{\mathrm{RTS}}_{k|N}\right]_{\theta\theta}.
\label{eq:rts-phase-variance}
\end{equation}
Here $\bm P^{\mathrm{RTS}}_{k|N}$ is the smoothed state covariance, and $N_p$ is the number of accepted pilot samples.  The model-based posterior diagnostic gave $V_{\mathrm{est}}=1.542\times10^{-3}$~rad$^2$. Combining this value with Eq.~\eqref{eq:linewidth-drift} gives $\sigma_\phi^2=V_{\mathrm{est}}+V_{\mathrm{drift}}=1.545\times10^{-3}$~rad$^2$ and $\xi_{\mathrm{RPN}}=1.046\times10^{-2}$~SNU at $V_A=6.77$ through Eq.~\eqref{eq:rpn-fitted-reference}.

\textbf{Transmitter preparation noise.} Let $\bm x_k$ be Alice's intended displacement, with $\mathbb E(\bm x_k\bm x_k^{\mathsf T})=V_A\mathbb{I}_2$.  After static calibration, write the residual displacement as $\bm e_k=\bm M_k\bm x_k+\bm d_k$.  The zero-mean residuals $\bm M_k$ and $\bm d_k$ are taken to be independent of $\bm x_k$.  The per-quadrature preparation noise is then
\begin{equation}
\xi_{\mathrm{mod}}
=
\frac{1}{2}\Tr\!\left[
V_A\,\mathbb E\!\left(\bm M_k\bm M_k^{\mathsf T}\right)
+\mathbb E\!\left(\bm d_k\bm d_k^{\mathsf T}\right)
\right].
\label{eq:mod-excess}
\end{equation}
Deterministic gain, rotation, and displacement are removed by calibration or absorbed into the fitted covariance.  Residual bias, extinction, and in-phase/quadrature (IQ) imbalance enter through $\bm M_k$ and $\bm d_k$.  The same matrices also carry DAC, electrical-drive, and thermal fluctuations. For one calibrated quadrature of the nested modulator, write $q'_k=\tilde\alpha_k\cos\varphi_k/2$ and $\varphi_k=\pi gU_{\mathrm{DAC},k}/U_\pi$.  Holding $\tilde\alpha_k$ fixed, the first-order drive and thermal error is
\begin{equation}
\delta q'_k
=
-\frac{\tilde\alpha_k}{2}\sin\varphi_k
\delta\varphi_k
+\mathcal O(\delta\varphi_k^2),
\qquad
\delta\varphi_k=
\frac{\pi g}{U_\pi}\,\delta U_{\mathrm{DAC},k}
+\frac{2\pi\beta_T L_{\mathrm h}}{\lambda_0}\,\Delta T_k
.
\label{eq:mod-quadrature-linearization}
\end{equation}
Here $g$ is the electrical gain, and $U_\pi$ is the half-wave voltage.  The effective thermo-optic coefficient is $\beta_T$, and $L_{\mathrm h}$ is the heated length~\cite{Padmaraju2013ThermalChallengesMRR}. Define $a_U=\pi g/U_\pi$ and $a_T=2\pi\beta_TL_{\mathrm h}/\lambda_0$.  The phase-error variance retains the covariance between drive and thermal fluctuations,
\begin{equation}
\operatorname{Var}(\delta\varphi_k)
=a_U^2\operatorname{Var}(\delta U_{\mathrm{DAC},k})
+a_T^2\operatorname{Var}(\Delta T_k)
+2a_Ua_T\operatorname{Cov}(\delta U_{\mathrm{DAC},k},\Delta T_k).
\label{eq:mod-phase-variance}
\end{equation}
Averaging Eq.~\eqref{eq:mod-quadrature-linearization} over accepted symbols supplies the covariance terms in Eq.~\eqref{eq:mod-excess}.

The measured electrical SNR fixes $a_U^2\operatorname{Var}(\delta U_{\mathrm{DAC},k})\simeq1/\mathrm{SNR}_{\mathrm{drive}}$.  The packaged-module value $\mathrm{SNR}_{\mathrm{drive}}\simeq55$~dB gives $\xi_{\mathrm{drive}}=2.14\times10^{-5}$~SNU at $V_A=6.77$.  The corresponding DAC floor is $1.58\times10^{-9}$~SNU.  The measured sub-kelvin per-symbol thermal excursion leaves the full modulation contribution at the same reported scale, giving $\xi_{\mathrm{mod}}\simeq2.1\times10^{-5}$~SNU.

\section{Shot-noise reference stability}\label{sec:app-shotnoise-allan}
\begin{figure}[htbp]
    \centering
    \includegraphics[width=0.5\linewidth]{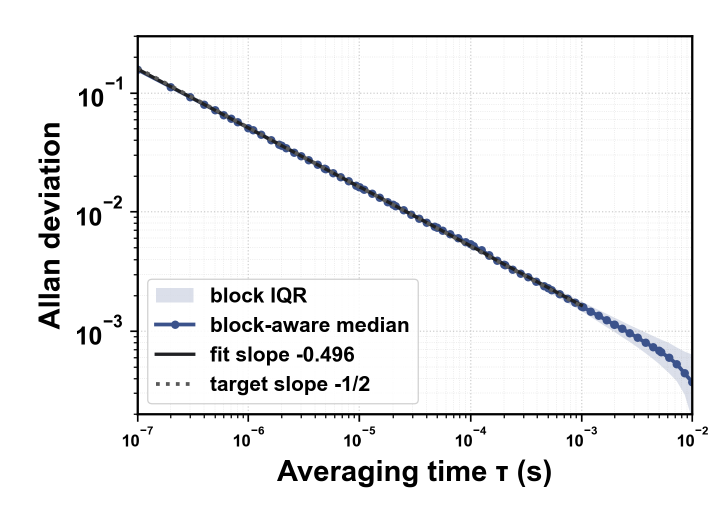}
    \caption{\textbf{Allan deviation of the fractional vacuum-noise power.} }
    \label{fig:app-shotnoise-allan}
\end{figure}

The vacuum-noise variance sets the SNU used to estimate the covariance matrix, transmittance, and excess noise.  A change in this reference between the quantum and vacuum acquisitions can bias these estimates~\cite{Ricard2025NoiseDynamics}.  As described in Section~4.3, each quantum window was therefore normalised by an adjacent vacuum acquisition under the same receiver settings. Figure~\ref{fig:app-shotnoise-allan} reports the Allan deviation of vacuum records acquired with these settings.  The data comprise $10^{4}$ acquisitions of $20$~ms each at $5\times10^{8}$~Sa/s, giving $200$~s of cumulative vacuum data.  Each acquisition was divided into non-overlapping blocks of $50$ samples ($0.1~\mu$s), and the shot-noise power was calculated for each block.  The Allan deviation was evaluated separately within each $20$-ms acquisition and summarized by the median and interquartile range across acquisitions.  Between $0.1~\mu$s and $1$~ms, the median fractional Allan deviation follows $\sigma(\tau)\propto\tau^{-1/2}$ with a fitted slope of $-0.496$, decreasing from $0.157$ to $1.6\times10^{-3}$. Because a single raw frame spans $0.2$~ms ($10^{5}$ ADC samples), this stable $\tau^{-1/2}$ scaling strictly encompasses the entire frame duration.  Across the $10^{4}$ acquisitions, the mean shot-noise power has a fractional standard deviation of $0.29\%$.

\section{Integrated platform}\label{sec:app-transceiver}

\subsection{Transmitter and receiver characteristics}\label{sec:app-bandwidth}

The transmitter and receiver dies described in Section~2.1 use standard silicon-photonic device structures on a silicon-on-insulator platform. The transmitter integrates carrier-depletion travelling-wave Mach--Zehnder modulators, thermal phase shifters, and edge couplers, whereas the receiver integrates $90^\circ$ optical hybrids, balanced germanium-on-silicon photodiodes, thermal phase shifters, and edge couplers. The transmitter and receiver dies are co-packaged with a quad-channel modulator driver and a transimpedance amplifier (TIA), respectively. The packaged modules were characterized through measurements of their small-signal electro-optic and optical-electrical responses, together with device-linearity tests. The dashed curves in Fig.~1(h) show the terminated-transmission-line model of the modulator.

At the component level, the responses were measured with a 67~GHz lightwave component analyzer.  As shown in Fig.~1(h), the silicon modulator 3-dB bandwidth grows with reverse bias from 0 to 4~V, which follows from the reduced junction capacitance of the carrier-depletion regime, and the 4.2~V driver supply holds the modulator in this high-bandwidth region.  The simulated electro-optic response of the transmission-line model, shown as the dashed curves in Fig.~1(h), follows the measurement across the 0 to 2~V range.  The integrated photodiode shows the same bias dependence and reaches its best response at 2~V, as shown in Fig.~1(j).

\begin{figure}[htbp]
    \centering
    \begin{subfigure}[b]{0.48\linewidth}
        \centering
        \panellabel{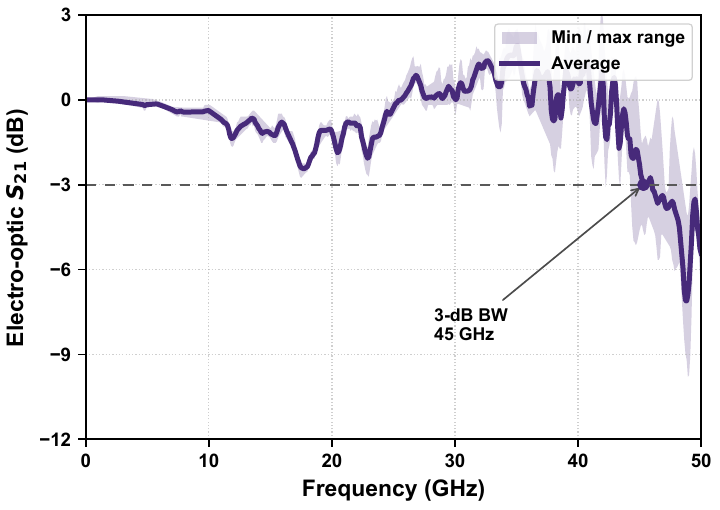}{(a)}
    \end{subfigure}
    \hfill
    \begin{subfigure}[b]{0.48\linewidth}
        \centering
        \panellabel{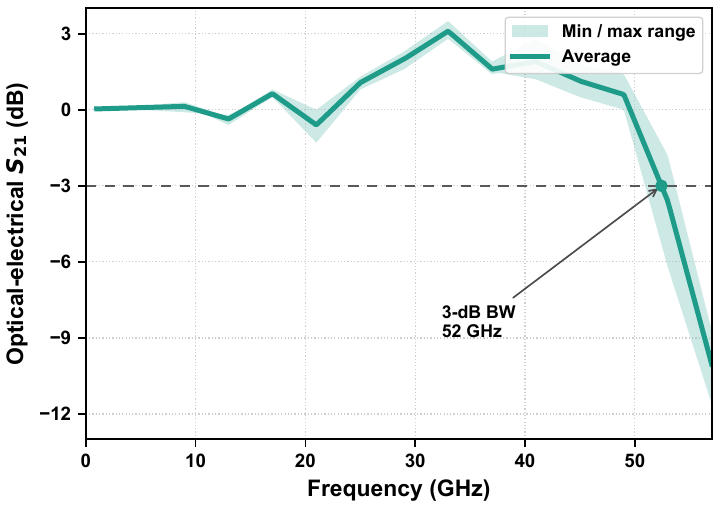}{(b)}
    \end{subfigure}
    \caption{\textbf{System-level frequency response of the bare transceiver die.} (a) End-to-end electro-optic response and (b) end-to-end optical-electrical response measured on the unpackaged chip.}
    \label{fig:transceiver_combined}
\end{figure}

At the system level, measured on the bare transceiver die, the end-to-end electro-optic response has an average 3-dB bandwidth near 45~GHz (Fig.~\ref{fig:transceiver_combined}(a)), and the optical-electrical response of the coherent receiver front end reaches about 52~GHz (Fig.~\ref{fig:transceiver_combined}(b)).  These bare-die bandwidths exceed the packaged-module values of 30 and 35~GHz quoted in Section~2.1, the difference being the parasitic roll-off added by the wirebonds and ball-grid-array transitions of the package, and both remain far above the 100~MBaud protocol symbol rate.

Multi-level modulation needs a linear drive chain.  We measured the output 1-dB compression point of the driver and modulator chain at 3.873~V peak-to-peak differential, so the modulator stays inside its linear range over the drive amplitudes used for state preparation.
\subsection{Self-injection-locking laser design}\label{sec:app-sil}
The silicon-nitride feedback chip was manufactured at the Shanghai Institute of Microsystem and Information Technology (SITRI) through a standard multi-project wafer (MPW) run. The 400-nm-thick silicon-nitride core is covered by a 3.8~$\mu$m top cladding and a 3.17~$\mu$m bottom cladding, which provides strong optical confinement while remaining compatible with standard processing. The chip integrates a spot size converter (SSC), a multimode interferometer (MMI), and a high-Q spiral microring resonator (MRR) embedded within a Sagnac loop configuration. A commercial DFB laser is butt-coupled to this feedback chip to form a laser. A high-reflection (HR) coating deposited on the rear facet of the DFB forms one mirror of the laser cavity, whereas the optical loop consisting of the MMI, the MRR, and the linking waveguides acts as the other. The SSC was designed to reduce inter-chip coupling loss, yielding a measured insertion loss below 2.5~dB. In the add-drop configuration, the input and drop ports of the MRR are connected through the MMI, so that the resonator operates as a wavelength-selective filter.
SIL state is established by adjusting the pump current of the DFB laser until its lasing wavelength coincides with the central resonance of the MRR. In the SIL state, the ratio of linewidth narrowing scales with the inverse square of the MRR $Q$ factor, which motivates a low-loss waveguide design. Widening the waveguide into the multimode regime reduces the modal overlap with the waveguide sidewalls, where surface roughness would otherwise introduce losses. The resulting propagation loss is measured to be below 5~dB/m, low enough to support a high $Q$ factor. The resonator employs Euler bends with a maximum radius of $R_{\max}=900~\mu$m and a minimum radius of $R_{\min}=240~\mu$m. Since the curvature of an Euler bend varies linearly along the propagation direction, the bend provides a gradual transition between straight and curved sections, preserving the adiabatic evolution of the fundamental mode, suppressing the excitation of higher-order transverse modes, and keeping the device footprint compact. The MRR has a perimeter of 87~mm and a coupling coefficient of 0.02, a combination chosen to obtain a high loaded $Q$ while keeping the insertion loss relatively low.

The ultra-high-$Q$ spiral was characterised through its drop-port transmission spectrum. The measured free spectral range (FSR) is approximately 0.014~nm, corresponding to 1.75~GHz, and the full width at half maximum (FWHM) is about 1.1~pm, from which a loaded-$Q$ factor of $1.4\times10^{6}$ is obtained.

\subsection{Automatic bias control}\label{sec:app-abc}

\begin{figure}[htbp]
    \centering
    \includegraphics[width=0.92\linewidth]{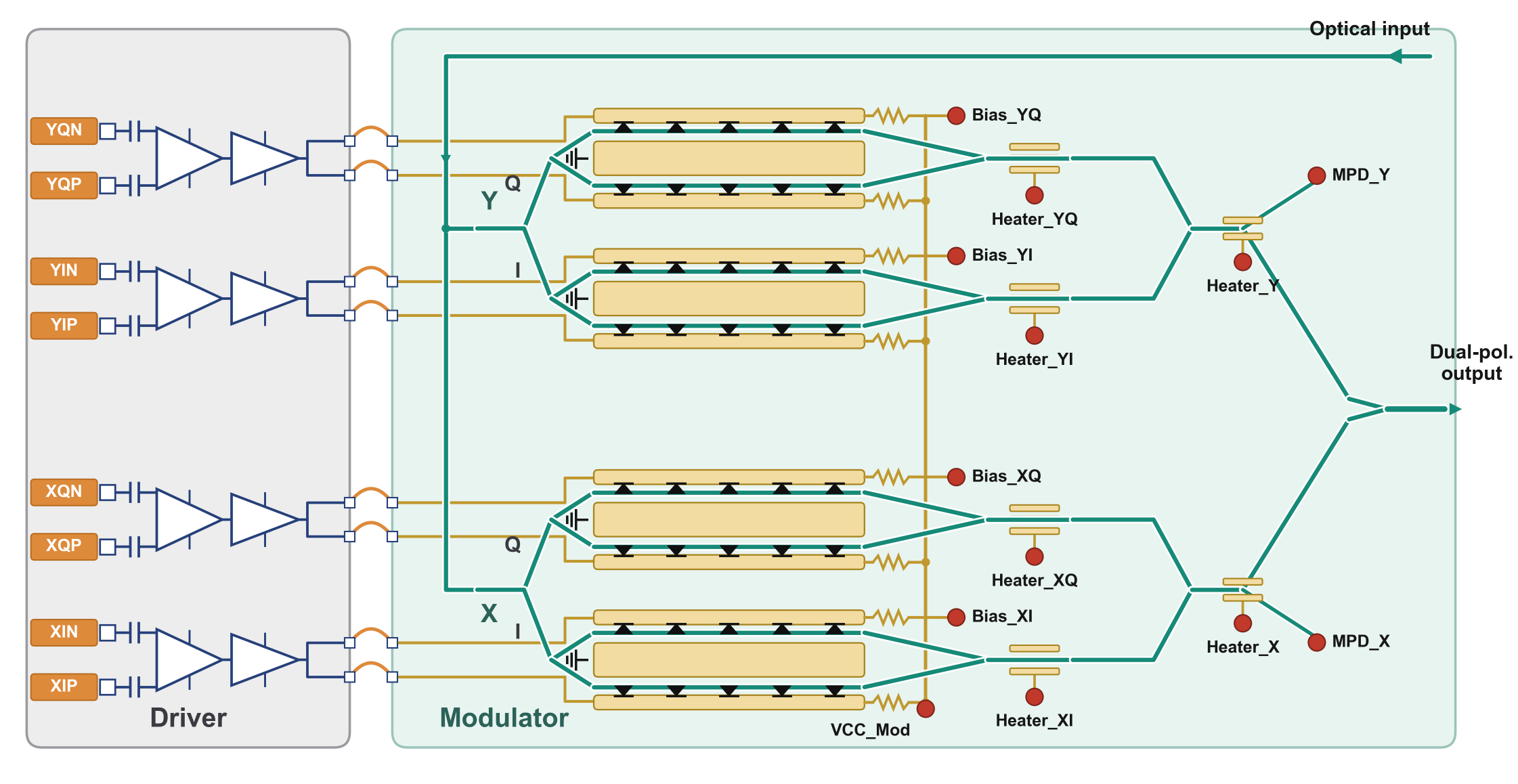}
    \caption{\textbf{Automatic bias control loop.}}
    \label{fig:dp-iqm-abc}
\end{figure}

A digital pilot-tone automatic bias control loop holds the transmitter stable over long runs.  A low-frequency pilot tone is added to the radio-frequency data signal and its optical response is monitored on the on-chip monitor photodiodes.  A field-programmable gate array (FPGA) digital lock-in demodulates the error signal and drives the thermal phase shifters through a proportional-integral controller, which compensates thermal drift and holds the modulators at their quadrature points with an extinction ratio above 22~dB over extended operation. The loop acts on the dual-polarization IQ modulator of Fig.~\ref{fig:dp-iqm-abc}, holding each nested Mach--Zehnder modulator at its transmission null and the in-phase and quadrature branches at a $\pi/2$ difference.

\section{Data acquisition}\label{sec:data-acq}

Quadrature waveforms were acquired using a Teledyne LeCroy 820Zi-B oscilloscope ($f_s = 5\times10^{8}$~Sa/s, $20$~ms/capture, $10^{7}$ ADC samples/channel) via a persistent VICP session to eliminate control overhead. To ensure strict I/Q synchronization, channels C1 and C2 were read simultaneously from frozen acquisition buffers under identical trigger events. Acquisition was decoupled from disk I/O using a multi-threaded producer--consumer queue with back-pressure, ensuring lossless, zero-frame-drop data logging. Raw payloads were stored as unscaled 8-bit integers alongside gain and offset metadata, allowing exact waveform reconstruction without memory-intensive float64 conversions.

\clearpage
\renewcommand{\refname}{Supplementary References}
\bibliographystyle{naturemag-doi}
\bibliography{ref}